\documentclass[12pt]{article}

\usepackage{newtxtext,newtxmath}

\usepackage{graphicx}
\usepackage{color}
\usepackage{booktabs}    
\usepackage{graphicx}    
\usepackage{array}       
\usepackage{amsmath}     
\usepackage{comment}
\usepackage{bm}

\usepackage[letterpaper,margin=1in]{geometry}

\renewenvironment{abstract}
{\quotation}
{\endquotation}

\date{}

\makeatletter
\renewcommand{\fnum@figure}{\textbf{Figure \thefigure}}
\renewcommand{\fnum@table}{\textbf{Table \thetable}}
\makeatother

\usepackage{scicite}

\usepackage{url}

\usepackage{bm}
\usepackage{amsmath}

\begin{document}



\def\scititle{
\LARGE {3D aperture-engineered diffractive neural networks \\ for super-resolution electromagnetic wave computing}
}
\title{\bfseries \boldmath \scititle}


\author{
\normalsize Sheng Gao$^{1}$, Songtao Yang$^{1}$, Haiou Zhang$^{1}$, Yuan Shen$^{1,2,\ast}$, and Xing Lin$^{1,2,\ast}$\\
\normalsize $^{1}$Department of Electronic Engineering, Tsinghua University, Beijing, 100084, China.\\
\normalsize $^{2}$Beijing National Research Center for Information Science and Technology, \\
\normalsize Tsinghua University, Beijing, 100084, China.\\
\normalsize $^\ast$Corresponding authors. Email: lin-x@tsinghua.edu.cn, shenyuan$\_$ee@tsinghua.edu.cn\\
}

\maketitle




\begin{abstract} \bfseries \boldmath

The rapid progress in 6G communication and high-bandwidth radar has driven an unprecedented surge in the spatial density of signal sources, resulting in an increasingly congested electromagnetic (EM) environment. When resolving closely spaced signals and interference, existing architectures are strictly bounded by the inherent diffraction limits of two-dimensional (2D) physical apertures, hindering super-resolution sensing and multi-interference mitigation in complex scenarios. Here, we present a 3D aperture-engineered diffractive neural network (AE-DNN) that achieves super-resolution sensing and computing by extending the traditional 2D aperture into 3D. The 3D aperture engineering framework is realized by constructing deep cascaded metasurface layers so that the diffractive propagation from oblique incident fields can be layer-wise modulated and piecewise encoded for perceiving EM fields far exceeding physical aperture limits. The $N$-layer AE-DNN has the capability to achieve ${\sim}N$ times higher angular resolution than the 2D aperture diffraction limit.The multi-dimensional synthetic aperture (MSA) training is developed to achieve speed-of-light coherent synthesis of the 3D aperture and integrate neural network-based modeling of multi-dimensional metasurface modulation. By orthogonalizing array response vectors in the analog domain, AE-DNN performs parallel super-resolution angle estimation, source number estimation, and source separation for up to 10 independent coherent or incoherent sources. Experimental results across the 36$\sim$41 GHz band demonstrate that AE-DNN resolves and suppresses closely spaced multi-interference by $\sim$20 dB, enhances communication capacity by 13.5$\times$, and reduces latency by three orders of magnitude. AE-DNN heralds a paradigm shift in signal processing for advanced radar and 6G communications.

\end{abstract}


\noindent

\subsection*{Introduction}

\begin{figure}[t!]
	\centering
	\includegraphics[width=0.95\textwidth]{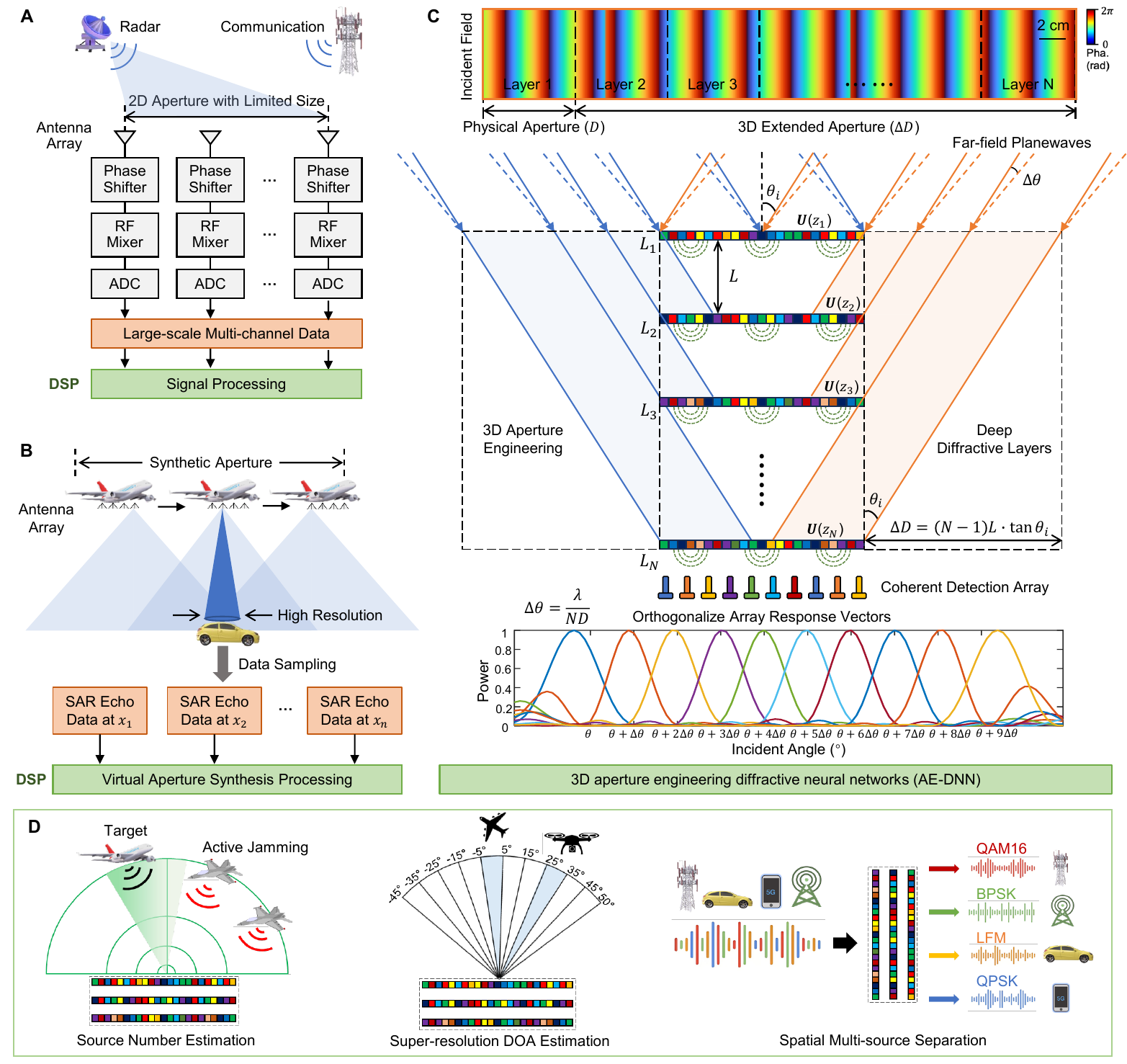}
	\caption{\textbf{AE-DNN using 3D aperture engineering.} (\textbf{A}) The traditional electronic architecture comprises an antenna array with a limited 2D aperture , phase shifters, RF mixers, ADCs, and DSPs, which leads to high processing complexity. (\textbf{B}) SAR enlarges the aperture size and enhances the resolution through time-consuming mechanical scanning and virtual aperture synthesis processing. (\textbf{C}) AE-DNN architecture utilizes deep cascaded metasurfaces to extend the perception aperture size with oblique incident waves and directly perform layer-wise modulation and coherent synthesis on the 3D extended aperture in the analog domain, thereby achieving array response vector orthogonalization and super-resolution sensing. (\textbf{D}) AE-DNN performs sensing and computing tasks at the speed of light, with high parallelism and low power consumption.}
	\label{fig_1}
\end{figure}

The rapid evolution of 6G wireless communications, high-bandwidth radar, and autonomous driving has driven an unprecedented surge in the spatial density of signal sources, pushing the electromagnetic environment toward extreme congestion.\cite{javadiRadarNetworksReview2020,dangWhatShould6G2020,liuEdgeComputingAutonomous2019}. Resolving closely spaced signals and interference with high resolution has become a formidable challenge for traditional digital architectures, where the spatial overlap of wavefields often leads to aliasing that is difficult to disentangle\cite{9149671,9394593}. Conventional paradigms rely on antenna arrays paired with phase shifters, RF mixers, and analog-to-digital converters (ADCs), where the digital signal processing (DSP) modules are required to process the captured large-scale multi-channel data stream to execute signal processing tasks \cite{krimTwoDecadesArray1996a,munozEnhancingFibreopticDistributed2022,wangOverviewEnhancedMassive2019,devosLOFARTelescopeSystem2009,wangOnchipTopologicalBeamformer2024}, resulting in high latency, power consumption, and cost (Fig.~\ref{fig_1}A). More fundamentally, these systems are physically bounded by the 2D apertures of their antenna arrays, strictly limiting their spatial resolution within the classical Rayleigh diffraction limit. Although synthetic aperture radar (SAR) can enhances the system resolution by synthesizing an larger virtual aperture (Fig.~\ref{fig_1}B), the time-consuming physical scanning of antenna platform and complex electronic computing fundamentally limit the real-time high-resolution sensing \cite{tanDirectionArrivalEstimation2014,moreiraTutorialSyntheticAperture2013}. Therefore, there is an urgent demand for a new generation of high-resolution, low-latency, and energy-efficient computational architectures. \cite{linArtificialIntelligenceBuilt2022,wangParallelInmemoryWireless2023,huangRadiofrequencySignalProcessing2025,wangSimplifiedRadarArchitecture2025}.

EM wave computing and photonic computing offers a promising path to executing tasks at the speed of light with low power and high parallelism, by utilizing EM fields rather than electrons as a computing carrier \cite{wetzsteinInferenceArtificialIntelligence2020a,shastriPhotonicsArtificialIntelligence2021}. On the one hand, the integrated photonic processors have demonstrated the capability to implement the mathematical integration and differentiation, blind source separation, and communication channel optimization \cite{shenDeepLearningCoherent2017,xu202111,feldmann2021parallel,pai2023experimentally,sludds2022delocalized,bogaertsProgrammablePhotonicCircuits2020,fengIntegratedLithiumNiobate2024,xuLargescalePhotonicChiplet2024,dongPartialCoherenceEnhances2024,milanizadehSeparatingArbitraryFreespace2022,zhangSystemonchipMicrowavePhotonic2024,seyedinnavadehDeterminingOptimalCommunication2023}. On the other hand, the diffractive EM wave computing in free-space perform the computations by modulating the multi-dimensional EM waves that can achieve object recognition, encrypted transmission, and communication encoding and decoding \cite{linAllopticalMachineLearning2018a,zhouLargescaleNeuromorphicOptoelectronic2021,chenAllanalogPhotoelectronicChip2023,gaoSuperresolutionDiffractiveNeural2024,yuAllopticalImageTransportation2025}. In contrast to integrated photonic circuits, which are constrained by optoelectronic conversion and on-chip scalability \cite{zhangSystemonchipMicrowavePhotonic2024,seyedinnavadehDeterminingOptimalCommunication2023}, spatial diffractive EM wave computing leverages metasurfaces\cite{yu2011light} with sub-wavelength structures to construct neural network architectures, enabling larger computing scale and facilitating more advanced computational tasks \cite{qianDynamicRecognitionMirage2022,liuProgrammableDiffractiveDeep2022,gaoSuperresolutionDiffractiveNeural2024}.

Despite these advances, the existing spatial EM wave computing architecture still confront fundamental  challenges when applied to complex EM environments. First, they are inherently limited by the diffraction limit of 2D aperture. Specifically, the physical aperture of the first modulation layer restricts the captured EM field, hindering the further improvement in system resolution and making it difficult to achieve super-resolution perception of closely spaced sources \cite{liuProgrammableDiffractiveDeep2022}. Second, current systems typically rely on electronic processing to perform angle estimation\cite{zhu2025selfcontrolleda}, yet they still fail to transcend the diffraction-limited resolution. Furthermore, because these systems are limited to the angle estimation task, they are incapable of the multi-source separation and coherent interference suppression required in complex EM scenarios.

To overcome these challenges, we propose a 3D aperture-engineered diffractive neural networks (AE-DNN) to achieve super-resolution EM sensing and computing in complex EM environments (Fig.~\ref{fig_1}C). By evolving from traditional 2D architectures to the 3D aperture framework, AE-DNN extends the system aperture by utilizing oblique incident fields with layer-wise diffractive modulation to process EM fields beyond the physical aperture. The novel 3D aperture engineering framework with deep diffractive layers substantially enhance its performance for super-resolution sensing and computational tasks. 
To achieve accurate modeling and broadband operational capability, we introduce the multi-dimensional synthetic aperture (MSA) training method, which enables the light-speed coherent synthesis of broadband EM waves captured by the 3D aperture and integrates neural network-based modeling of multi-dimensional metasurface modulation into a unified optimization framework. 
By orthogonalizing array response vectors in the analog domain, AE-DNN executes parallelized super-resolution direction-of-arrival (DOA) estimation, source number estimation, and high-isolation source separation for up to 10 independent sources at the speed of light (Fig.~\ref{fig_1}D). Leveraging metasurface focusing gain to amplify signal power by 13 dB, AE-DNN becomes broadly applicable to complex EM environments, including radar anti-jamming and integrated sensing, communication, and computing applications.

\subsection*{3D aperture engineering framework}

The major advantage of AE-DNN over existing diffractive neural networks is the 3D aperture engineering framework, which allows the scalable super-resolution with additive layers to surpass the classical diffraction limit imposed by the single-layer 2D physical aperture size and improve the detection energy efficiency and channel crosstalk suppression ratio (CSR). By leveraging oblique-incident plane wave propagation, a deep cascade of $N$ metasurfaces, each of size $D$ comprising a dense array of modulation units with numbers of $K \times K$ and separated by a distance $L$, captures EM fields originating beyond the first layer's boundary (Fig.~\ref{fig_1}C). AE-DNN performs layer-wise coherent synthesis of these wavefronts via amplitude and phase modulation of polarized EM fields over a broadband frequency range, which substantially expands the effective receiving aperture for incident fields of both coherent and incoherent sources.

Modeling incident wavefront propagation with ideal geometric rays, the 3D extended system aperture size is given by $\Delta D = (N-1)L \cdot \tan{\theta}$, which leads to an effective aperture size of $D_{\mathrm{eff}} = D + \Delta D$. For bilateral incidence, the total effective aperture is $D_{\mathrm{eff}} = D + 2\Delta D$. Therefore, based on the Rayleigh criterion \cite{woodPhysicalOpticsRobert1919}, the angular resolution of AE-DNN is set and improved to:
\begin{equation}
\Delta \theta = {\lambda_c} \, / \, {D_{\mathrm{eff}}},
\end{equation}
where $\lambda_c$ represents the central wavelength. The maximum effective aperture size can be achieved by optimizing inter-layer spacing to $L = D/\tan{\theta}$, with which $D_{\mathrm{eff}} = ND$ and $D_{\mathrm{eff}} = (2N - 1) D$ for unilateral and bilateral incidence, respectively. With the extended effective aperture size, the corresponding number of modulation units is $K_1 \times K_1$, where $K_1 = K \, D_{\mathrm{eff}}/D$. To clarify the use of ``super-resolution'' in this study, we define the diffraction limit resolution as the classical Rayleigh diffraction limit angular resolution dictated by the physical aperture size of a single layer at the operating wavelength. Unlike SAR, our architecture offers light-speed super-resolution sensing and multi-channel source separation, capable of improving angular resolution by a factor of $N$ without mechanical scanning or complex digital post-processing.

\subsection*{Orthogonalize array response vectors in the analog domain}

Considering the array response to $n$ far-field channels, the wideband system output captured by $m$ ($m \leq n$) antenna detectors can be modeled as $\mathbf{y}(t) = \int_{B} \mathbf{G}(\bm{\theta}, f) \, \mathbf{S}(f) \, e^{j 2\pi f t} \, df + \mathbf{n}(t)$, assuming an operating frequency range of width $B$ centered at $f_c$. Here, $\mathbf{y}(t) \in \mathbb{C}^m$ is the received signal vector at time $t$; $\mathbf{S}(f) \in \mathbb{C}^n$ represents the source signal spectrum corresponding to the time-domain signal $\mathbf{s}(t) \in \mathbb{C}^n$; and $\mathbf{n}(t) \in \mathbb{C}^m$ denotes the noise vector \cite{krimTwoDecadesArray1996a}. The frequency-dependent array manifold matrix $\mathbf{G}(\bm{\theta}, f) = [\mathbf{a}(\theta_1, f), \dots, \mathbf{a}(\theta_n, f)] \in \mathbb{C}^{m \times n}$ consists of response vectors $\mathbf{a}(\theta_i, f)$ for $i = 1, \dots, n$, each corresponding to a channel from incident angle $\theta_i$. The incident angles $\bm{\theta} = [\theta_1, \theta_2, \dots, \theta_n]$ are uniformly spaced by the angular interval $\Delta \theta$. The electronic DOA estimation \cite{wangOverviewEnhancedMassive2019}, e.g., multiple signal classification (MUSIC), rely on eigenvalue decomposition using DSP that causes a high complexity of $O(m^3)$. Moreover, with $\Delta \theta$ approaching diffraction-limited resolution, the correlation between $\mathbf{a}(\theta_i,f)$ vectors sharply rises, causing rank deficiency, i.e., $\mathrm{rank}(\mathbf{G})\ll \min(m,n)$, and prevents the stable estimation of source signals $\mathbf{s}$.

To overcome these, AE-DNN physically optimizes channel performance with the deeply cascaded diffractive metasurfaces for pre-detection computing. The discretized detection regions at output plane are designed to directly receive the separated channel signals to reduce the row-vector correlation of $\mathbf{G}(\bm{\theta}, f)$. AE-DNN intrinsically achieves the transformation $\mathbf{O}(\bm{\theta}, f) \in \mathbb{C}^{n \times n} $ that performs the change of basis for the column space of $\mathbf{G}(\bm{\theta}, f)$ to orthogonalize array response vectors and construct independent source channels:
\begin{equation}
\mathbf{y}(t) = \int_{B} \mathbf{G}(\bm{\theta}, f) \, \mathbf{O}(\bm{\theta}, f) \, \mathbf{S}(f) \, e^{j 2\pi f t} \, df + \mathbf{n}(t) = \mathbf{\Sigma} \, \mathbf{s}(t) + \mathbf{n}(t).
\end{equation}
Here, $\mathbf{\Sigma} = \mathbf{G}(\bm{\theta}, f) \, \mathbf{O}(\bm{\theta}, f) \approx \begin{bmatrix} \mathbf{0}_{m \times k} & \mathbf{\tilde{\mathbf{\Sigma}}}_{m \times m} & \mathbf{0}_{m \times (n - m - k)} \end{bmatrix}$, with $0 \le k \le n-m$, represents the orthogonalized array response matrix. Specifically, the diagonal block $\mathbf{\tilde{\mathbf{\Sigma}}}_{m \times m} = \text{diag}(\tilde{\sigma}_{k+1}, \dots, \tilde{\sigma}_{k+m})$ denotes the beam focusing complex gains for the reconstructed wideband channels corresponding to angular range from ${\theta}_{k+1}$ to ${\theta}_{k+m}$. Therefore, training AE-DNN for orthogonalization implicitly aims at deriving the null-space matrix and the scaled right inverse of the latent array response $\mathbf{G}(\bm{\theta}, f)$, which is independent of incident signal waveform. The 3D extended aperture $D_{\mathrm{eff}}$, responding to the large-scale EM field at the incident plane, endows $\mathbf{O}(\bm{\theta}, f)$ with properties that stabilize the solution to the ill-conditioned problem. This ensures that $\mathrm{rank}(\mathbf{\Sigma}) = m$ and enables channel separation and suppression at the super-resolved angular resolution $\Delta \theta$ defined in Eq.~1, working for both coherent and incoherent sources.

Given the separated channels and known focusing power gains $|\tilde{\sigma}_{i}|^2$, $i = k+1, \dots, k+m$, the average power of the detected signal at each angle can be represented as $P_y(\theta_i) = |\tilde{\sigma}_{i}|^2 P_s(\theta_i)$, where $P_s(\theta_i)$ denotes the average power of the incident signal. Therefore, the number of sources is determined by applying a threshold to $P_y(\theta_i)$. The DOA estimation is implicitly performed via the indices of the suprathreshold $P_y(\theta_i)$ values, which correspond to the target angles. Furthermore, the beam focusing gains substantially enhance the signal-to-noise ratio (SNR) and maximizes the channel communication capacity:
\begin{equation}
C \approx \sum_{i=k+1}^{k+m} B_i \log_2 \left( 1 + \frac{P_y(\theta_i)}{P_{\text{noise}} + \sum_{j \neq i} I(\theta_j \to \theta_i)} \right)
\end{equation}
where $B_i, P_y(\theta_i)$, and $P_{\mathrm{noise}}$ denote the bandwidth, signal power, and noise power of each channel, respectively. $I(\theta_j \to \theta_i)$ quantifies the system crosstalk from the $j$-th source to the $i$-th detection region due to the imperfection of the orthogonalization defined in Eq.~2. Ultimately, AE-DNN replaces complex digital post-processing with passive, parallel EM wave computing, enabling super-resolution sensing with low power consumption.

\begin{figure}[t!]
	\centering
	\includegraphics[width=0.92\textwidth]{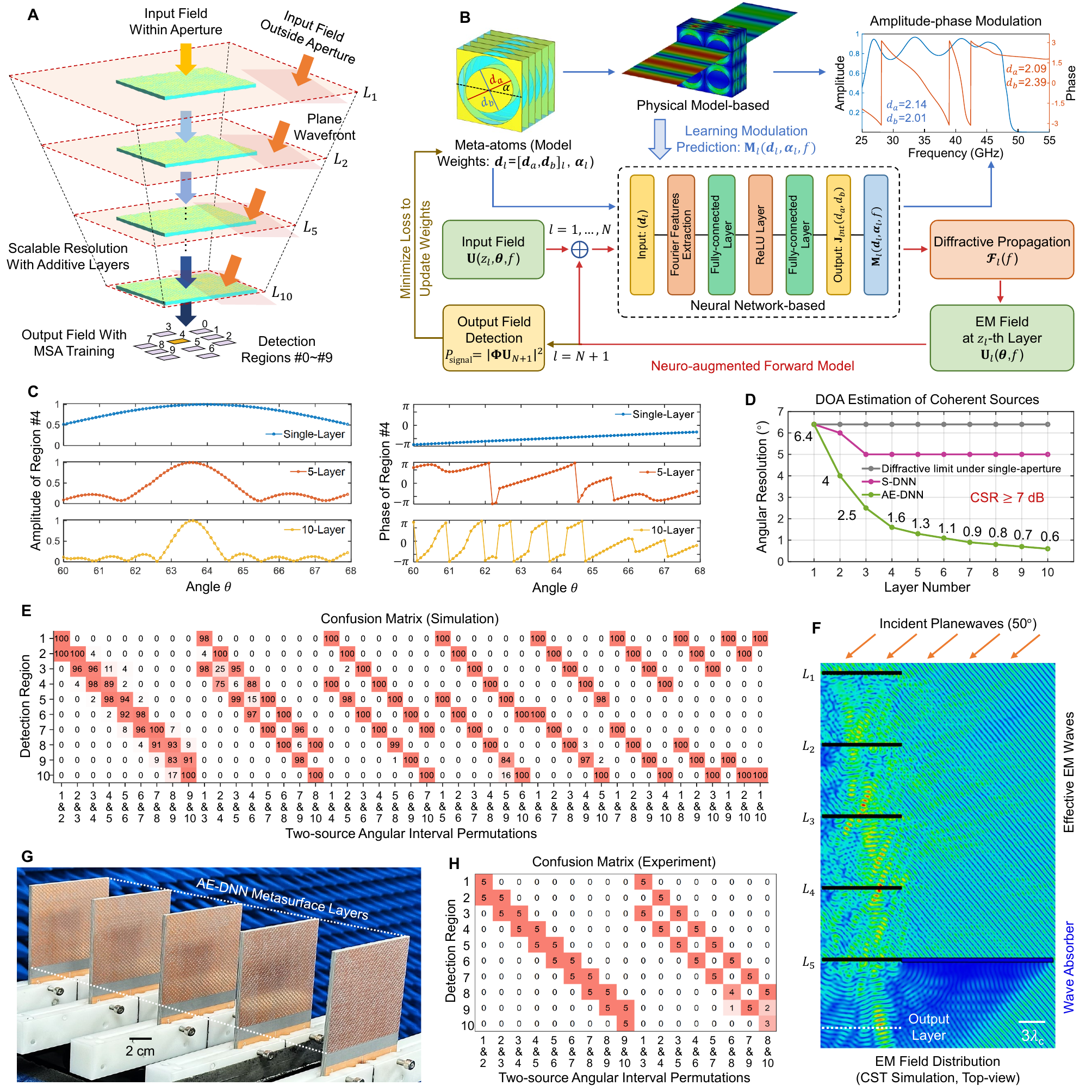}
	\caption{\textbf{MSA training of AE-DNN for super-resolution sensing.} (\textbf{A}) Both input fields within and outside apertures are utilized to train the model to achieve light-speed aperture synthesis for super-resolution sensing. (\textbf{B}) MSA training method is accelerated with the neuro-augmented forward model that integrates mini-FCNNs to characterize the high-dimensional modulation of meta-atoms. Meta-atom geometries are updated by minimizing the loss function during training. (\textbf{C}) As the number of layers increases, the phase of output fields exhibit higher-order nonlinear responses to the incident angle. (\textbf{D}) AE-DNN achieves scalable super-resolution for DOA estimation. Increasing the number of layers continuously improves the angular resolution while maintaining high energy efficiency and CSR. (\textbf{E}) Confusion matrix for two-target DOA estimation in simulation. (\textbf{F}) Full-wave simulation showing EM field distribution in the AE-DNN. (\textbf{G}) A five-layer AE-DNN comprising metasurface layers. (\textbf{H}) Experimental confusion matrix for two-target DOA estimation.}
	\label{fig_2}
\end{figure}

\subsection*{MSA training process}

The AE-DNN model is designed and evaluated through a deep learning-based optimization method. Based on 3D aperture, AE-DNN effectively processes the EM field within the physical aperture from previous layers and obliquely incident plane waves originating from beyond the physical aperture. AE-DNN further optimizes the model weights through MSA training, performs precise segmented modulation and light-speed coherent synthesis of the extended aperture, thereby achieving super-resolution sensing and computing (Fig.~\ref{fig_2}A). The linear-polarized incident EM field on the 3D extended aperture with a unit number of $K_1 \times K_1$ can be denoted as $\mathbf{U}(z_l, \bm{\theta}, f) \in \mathbb{C}^{K_1^2 \times n}$. Originating from a source signal with azimuth angle $\theta_i$ and constant elevation angle $\varphi_i$, the piecewise encoded EM field at the $l$-th layer, with an axial position $z_l$ ($l \in \{1, \dots, N\}$), is formulated as:
\begin{equation}
[\mathbf{U}(z_l, \bm{\theta}, f)]_{r,\,i} = \tilde{A} \exp\left\{ j \kappa \left( x_q \sin \varphi_i + y_p \cos \varphi_i \sin \theta_i \right) \right\},
\end{equation}
where $\tilde{A} = A \exp\{ j \kappa z_l \cos \varphi_i \cos \theta_i \}$ denotes the constant complex value at the $l$-th layer, while $A$ and $\kappa = 2 \pi f / c$ denote the amplitude and wavenumber, respectively, with $c$ being the speed of light \cite{gaoSuperresolutionDiffractiveNeural2024}. Here, the linear index $r$ maps discretized coordinates $(x_q, y_p)$ via $r = (p-1)K_l + q$. For the first layer ($l=1$), $p$ and $q$ range from $1$ to $K$. For subsequent layers ($l > 1$), the coordinate boundaries shift linearly based on the spatial offsets $\Delta D_p = L \tan \theta_i$ and $\Delta D_q = L \tan \varphi_i$, scaled by the factor $K/D$. Specifically, the EM field at the $l$-th layer $\mathbf{U}_l(\bm{\theta}, f) \in \mathbb{C}^{K_1^2 \times n}$, comprising EM fields from the previous layer and obliquely incident plane waves, with initial state $\mathbf{U}_{0} = \mathbf{0}$, can be represented as:
\begin{equation}
\mathbf{U}_l(\bm{\theta}, f) = \bm{\mathcal{F}}_l(f) \, \text{diag}(\mathbf{M}_l({\bm{d}_l}, f)) \left( \mathbf{U}_{l-1}(\bm{\theta}, f)+\mathbf{U}({z_l}, \bm{\theta}, f) \right),
\end{equation}
where $\mathbf{M}_l({\bm{d}_l}, f) \in \mathbb{C}^{K_1^2 \times 1}$ denotes the metasurface modulation with $\bm{d}_l$ representing geometric parameters of meta-atoms, and $\bm{\mathcal{F}}_l(f) \in \mathbb{C}^{K_1^2 \times K_1^2}$ denotes the Rayleigh–Sommerfeld diffractive propagation \cite{linAllopticalMachineLearning2018a}. By adjusting the geometric parameters of the meta-atoms, the amplitude and phase of polarized EM waves can be independently controlled across a broad wavelength range. Thus, the total output EM field at the detection plane is given by:
\begin{equation}
\mathbf{U}_{N+1}(\bm{\theta}, f) = \sum_{w = 1}^{N} \left( \left( \prod_{l=w}^{N} \bm{\mathcal{F}}_l(f) \, \text{diag}(\mathbf{M}_l({\bm{d}_l}, f)) \right) \mathbf{U}({z_w}, \bm{\theta}, f) \right).
\end{equation}

The sampled field of detection regions $\mathbf{T}(\bm{\theta}, f) \in \mathbb{C}^{m \times n}$ responses at the output plane can be formulated as: $\mathbf{T}(\bm{\theta}, f) = \mathbf{\Phi} \, \mathbf{U}_{N+1}(\bm{\theta}, f)$, where $\mathbf{\Phi} \in \{0, 1\}^{m \times K_1^2}$ is a row-selection matrix that extracts $m$ observation points from the full spatial grid. Each row of $\mathbf{\Phi}$ corresponds to a standard basis vector for selecting a specific measurement channel. Therefore, the loss function of MSA training for optimizing meta-atom geometric parameters is given by:
\begin{equation}
\min_{\mathbf{d}_l} \left\| |\mathbf{T}(\bm{\theta}, f)|^2 - |\mathbf{\Sigma}|^2 \right\|_F^2 + \lambda \left\| |\mathbf{U}_{N+1}(\bm{\theta}, f) \mathbf{1}_n|^2 - \mathbf{\Phi}^T \mathbf{1}_m \right\|_2^2
\end{equation}
where the first loss term enforces data fidelity within the detection regions, while the second loss term maximizes the total detected power, weighted by the regularization parameter $\lambda$. The azimuth angle is varied during the learning process to enhance the model's robustness against angular changes. During the training, the miniature fully connected neural network (mini-FCNN) is designed to learn the multi-dimensional modulation of each meta-atom based on the geometric parameters $\mathbf{d}_l = [\mathbf{d}_a, \mathbf{d}_b]_l$, which enables precise modeling of metasurface modulation (see Fig.~\ref{fig_2}B). The mini-FCNN is trained on a CST Studio Suite generated database to model the complex EM-response of meta-atoms, serving as a differentiable operator in the forward model (supplementary text 1.1). The resulting neuro-augmented physical model, integrated with the mini-FCNN, effectively minimizes the loss function to accurately update the model weights layer by layer. The MSA training unifies the physical model optimization and device inverse design process, featuring multi-dimensional signal processing and high-model accuracy and robustness.

\subsection*{Scalable super-resolution sensing}

Based on the 3D aperture engineering framework, AE-DNN supports scalable super-resolution sensing, meaning that a $N$-layer AE-DNN has the capability to achieve ${\sim}N$-fold angular resolution beyond the single-aperture diffraction limit. Since the incident angle is nonlinearly embedded in the polarized incident EM field via the phase term $\sin{\theta_i}$ (Eq.~4), the piecewise encoding of the high-dimensional EM field across successive layers with 3D aperture engineering generates a complex nonlinear angular response. With MSA training, the higher-order nonlinear angular response can be obtained as the depth of architecture increases. Fig.~\ref{fig_2}C plots the amplitude and phase responses of the fourth detection region with respect to $\theta$ for three AE-DNNs with 1, 5, and 10 metasurface layers. Each layer comprises $23 \times 23$ meta-atoms with a unit size of $3\,mm$, resulting in a physical aperture size of $D = 8.97\lambda_c$ by setting central frequency $f_c = 39\,\text{GHz}$. The inter-layer distance is set to $L = 8\lambda_c$. The angular response verifies that increasing the layer number expands the effective aperture for piecewise deep encoding of incident EM waves, narrowing the focusing beam while enabling higher-order phase nonlinearity. This enhanced nonlinear angular response indicates a sharpened sensitivity of the system to variations in the incident angle.

Based on this nonlinear sensing mechanism and 3D aperture engineering framework to process larger effective EM wave regions, AE-DNN supports scalable super-resolution sensing for DOA estimation of coherent sources with additive layers (Fig.~\ref{fig_2}D). By setting CSR above 7~dB to facilitate robust detection during MSA training, increasing the AE-DNN layers from 1 to 10 continuously improves angular resolution from a diffraction-limited $6.4^\circ$ to a super-resolved $0.6^\circ$, with the initial angular range initialized at $50^\circ$. This performance also surpasses the single-aperture Rayleigh diffraction limit by a factor of $\sim$10.7 times and outperforms super-resolution diffraction neural networks (S-DNN) \cite{gaoSuperresolutionDiffractiveNeural2024}. The results verify that the $N$-layer AE-DNN has the capability to achieve ${\sim}N$-fold angular resolution beyond the single-aperture diffraction limit. Further numerical evaluations of 5-layer AE-DNN on varying meta-atoms at each layer from $23 \times 23$ to $92 \times 92$ also show the angular resolution 3$\sim$5 times higher than S-DNN (supplementary text 2.1). AE-DNN maintains the detection energy efficiency above 1$\%$, even as the architecture scales beyond five layers.

To enable the experimental validation with error tolerance, we designed a 5-layer AE-DNN with $D = 8.97\lambda_c$ and $L = 8\lambda$ and set an angular resolution to $1.6^\circ$ with an angular range from $50^\circ$ to $66^\circ$. Ten detection regions are utilized to map ten angular intervals, with the source numbers and intervals determined by identifying the top-$n$ detection regions with the highest power. The designed processor fully exploits the effective EM wave outside the aperture with the layer-wise metasurfaces within the entire angular range, representing a four-fold increase in angular resolution compared to the single-aperture diffraction limit. Considering the complete sensing field of view, the transverse baseline length of the input plane wave aperture was configured as $8D$ during the training. The five deeply cascaded metasurfaces of AE-DNN is fabricated using high-precision standard PCB technology, with each meta-atoms providing a sufficient modulation bit depth of 5-bit for model quantization (detailed in supplementary text 1.2).

\begin{figure}[t!]
	\centering
	\includegraphics[width=0.95\textwidth]{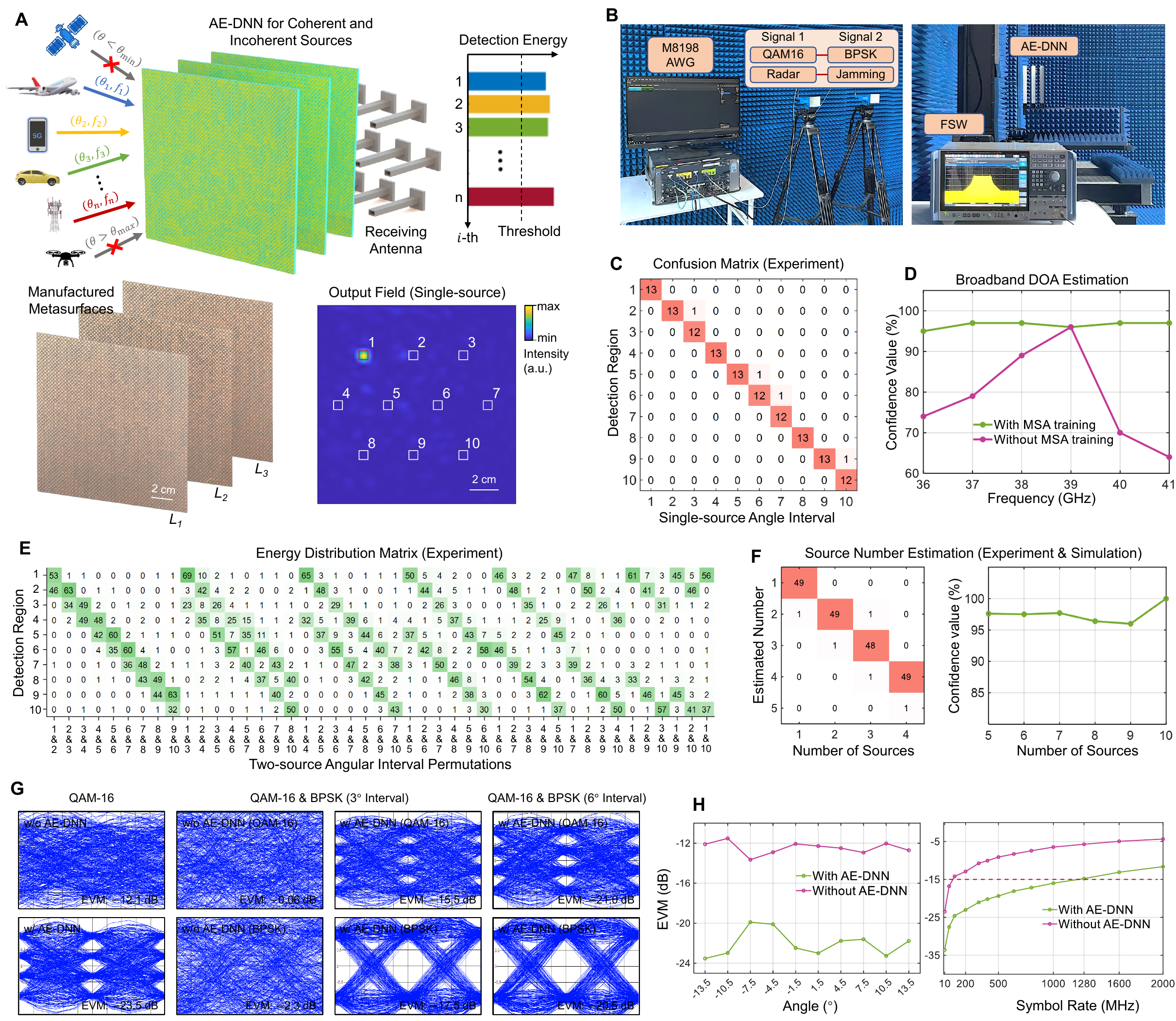}
	\caption{\textbf{AE-DNN for parallel sensing and computing tasks.} (\textbf{A}) AE-DNN selectively maps coherent and incoherent sources from different angles to the corresponding receiving antennas (top). The manufactured metasurfaces and a representative experimental output field of a single source is shown (bottom). (\textbf{B}) Experimental system utilizes AWG to generate independent source signals with FSW to analyze the processed signals. (\textbf{C}) Experimental confusion matrix evaluated on the single-source testing dataset. (\textbf{D}) MSA training enables the DOA estimation under broader wavelength range. (\textbf{E}) Experimental energy distribution matrix for 3$^\circ$ super-resolution DOA estimation of two sources under 45 angular interval permutations. (\textbf{F}) Experimental confusion matrix (left) and numerically evaluated confidence values (right) for source number estimation with $m$ varying from 1 to 10. (\textbf{G}) Eye diagrams and EVMs of QAM-16 and BPSK signals received by spectrum analyzer with and without AE-DNN. (\textbf{H}) EVMs of received QAM-16 source signals from different angles (left) and at different symbol rates (right). AE-DNN substantially improve the communication capacity.}
	\label{fig_3}
\end{figure}

With the MSA training to coherent synthesis of effective incident EM wavefront, the trained neuro-augmented forward model was numerically tested for super-resolution DOA estimation using a two-source dataset comprising 4500 samples at angular intervals under arbitrary incident angle permutations within the field of view. The confusion matrix in Fig.~\ref{fig_2}E illustrates the angular interval prediction for two coherent sources based on the power detected at each region, where the confidence value indicating the prediction accuracy reaches $96.9\%$. The successful estimation of two sources at angular intervals under arbitrary incident angle permutations indicates that the model can be generalized to estimate up to ten sources. Fig.~S3A shows the corresponding CST validated energy matrix for two-source super-resolution DOA estimation under 45 different angular interval permutations, achieving a confidence value of $97.7\%$. Fig.~\ref{fig_2}F shows the top-view of the CST full-wave EM field distribution in AE-DNN, illustrating an example of utilizing the extended aperture to receive effective incident EM waves for estimating the source angle at $50^\circ$. At the output layer, an additional absorbing material was used to eliminate interference from direct illumination (Eq.~6). The fabricated 5-layer AE-DNN is shown in Fig.~\ref{fig_2}G. To validate the system angular resolution in the experiment, two transmitting antennas were placed with angular separations of $1.6^\circ$ and $3.2^\circ$, and a total number of 85 test samples of two-source angular interval were measured by rotating the AE-DNN using a high-precision motorized rotary stage and analyzed with vector network analyzer (supplementary text 1.3). The confusion matrix (Fig.~\ref{fig_2}H) and energy matrix (Fig.~S3B) of experimental results validate that AE-DNN achieves the $1.6^\circ$ super-resolved angular resolution with a confidence value of $96.5\%$.

\subsection*{Parallel sensing and computing tasks}

By orthogonalizing array response vectors in the analog domain, the AE-DNN is capable of executing multiple sensing and computing tasks in parallel, and can be applied in complex EM interference environments. We designed and constructed a three-layer cascaded metasurface system, integrated with $m=10$ receiving antennas at the output layer (Fig.~\ref{fig_3}A, top). This three-layer AE-DNN is engineered to execute three pivotal tasks, including super-resolution DOA estimation, multi-source number estimation, and high-CSR source separation in the broad mmWave band (36–41 GHz) at the central wavelength of $f_c = 39$ GHz. Each metasurface consists of $45 \times 45$ meta-atoms, each with a size of $3$~mm, leading to a Rayleigh diffraction limit resolution of $3.3^\circ$. To balance the sensing resolution and CSR, AE-DNN is optimized with MSA training within an angular range of $-15^\circ$ to $15^\circ$ with ten angular intervals, achieving an angular super-resolution of $3^\circ$ across the frequency range from $36$ to $41$~GHz. When far-field EM waves from sources are incident at angles $ \theta_i \in \{-15^\circ+3(i-1),\, -15^\circ+3i\}, i=1,\cdots,n$, with $n=10$, AE-DNN orthogonalizes the array response vectors that maps the energy of each source signal to the $i$-th receiving antenna, while suppressing signals from other angles. The evaluation of AE-DNN with CST full-wave simulation validates the effectiveness of trained neuro-augmented model (supplementary text 2.2). The metasurfaces are prototyped using PCB manufacturing based on the exported 3D model as shown in Fig.~\ref{fig_3}A(bottom, left). During the experiment, a high-precision angular rotation stage is utilized to mount AE-DNN and evaluate its performance for processing source signals under different angular interval permutations (supplementary text 1.3). To demonstrate the capability of AE-DNN for parallel sensing and computing tasks, the experimental platform leverages arbitrary waveform generator (AWG) to generate independent source signals to far-field transmitting antennas and FSW spectrum analyzer to analyze the received signals from probe waveguide antennas (Fig.~\ref{fig_3}B).

We first demonstrate that the AE-DNN can orthogonalize the array response vectors of communication channels in the analog domain with the single-source experiments, i.e., $\mathbf{s}(t) = \mathbf{e}_i \in \mathbb{R}^{10}$, $i = 1,..., 10$, with $\mathbf{e}_i$ being the standard basis vector. Fig.~\ref{fig_3}A(bottom, right) shows an exemplar DOA estimation for an incident angle of $-13.5^\circ$, where the AE-DNN successfully maps the incident source energy to a focal spot within the first detection region, indicating the source angular interval of [$-15^\circ$, $-12^\circ$]. The highly matched output fields of neuro-augmented forward model, CST full-wave simulation, and experimental measurement validate the MSA training method for optimizing model geometric parameters robust to the system errors. The correct detection region accounted for $96\%$ high energy percentage of the total detected energy, with a channel crosstalk below $-19$~dB and a beam focusing gain of $12.4$~dB. Furthermore, using a $3^\circ$ step size, we measured the detection regions energy for ten incident angles $-13.5^\circ,-10.5^\circ,\cdots,13.5^\circ$ to construct the output energy response vector, i.e., $\mathbf{P}_y\in \mathbb{R}^{10}$, of AE-DNN (Table.~S1). The results indicate that when the incident angle is $\theta_i$, the $i$-th element of the energy response vector $P_y(\theta_i)$ is substantially larger than the others. On average, the main channel contained $87\%$ of the total energy, with a channel crosstalk remained below $-12.3$~dB, and the beam focusing gain reached $12$~dB. These experimental findings are in strong agreement with simulation results (Table.~S2), proving that the AE-DNN orthogonalizes the array response vectors and greatly enhances source signal power. Additionally, to comprehensively evaluate the single-source DOA estimation performance of the AE-DNN, we continuously sampled 130 angles within [$-15^\circ$, $15^\circ$] at $0.2^\circ$ step size (excluding the interval boundary). Experiments show the single-source DOA estimation classification accuracy of $96.0\%$, and the corresponding confusion matrix is presented in Fig.~\ref{fig_3}C. Besides, the AE-DNN achieved a confidence value above $95\%$ for DOA estimation over the broadband frequency range between $36$~GHz and $41$~GHz (Fig.~\ref{fig_3}D). This broadband capability stems from the MSA training method. Without it, AE-DNN performance would be confined to the training frequency of $39$~GHz and degrade substantially at other frequencies. The MSA training mitigates experimental errors due to metasurface modeling inaccuracies and material resonance frequency shifts, endowing the AE-DNN with superior broadband characteristics.

We further demonstrate that the AE-DNN achieves super-resolution DOA estimation with an angular resolution of $3^\circ$. In multi-source experiments, DOA estimation for $n$ source signals is achieved by identifying the $n$ highest energy values among the ten detection regions. Here, $n$ denotes the number of source signals in free space, determined via source number estimation. Since the AE-DNN has ten detection regions, it can estimate DOA for up to ten coherent sources from distinct angles in parallel. This experiment utilized two transmitting antennas with a spacing of $\Delta\theta$ to transmit two source signals that are readily scalable to multiple sources. Different angular interval permutations were conducted by rotating the stage, with $\Delta\theta$ ranging from $3^\circ$ to $30^\circ$. Fig.~\ref{fig_3}E shows the energy distribution matrix for super-resolution DOA estimation under 45 two-source angular permutations. Across the field of view of angular ranges, the AE-DNN achieved super-resolution DOA estimation for arbitrary two-source angular permutations with $100\%$ confidence, and an average energy percentage of $44\%$ in the correct detection regions.

The AE-DNN can perform source number estimation tasks for up to 10 sources. By setting an appropriate threshold, angular intervals corresponding to receiving antennas with energy above this threshold are identified as containing source signals. Due to the beam focusing gain across ten receiving antennas being non-uniform (Table.~S1), pre-calibration is required. After calibration, the detected energy at each antenna was normalized to facilitate comparison using a fixed threshold (empirically set to $0.3$). In experiments, a one-to-four power divider was used to connect the AWG to multiple far-field transmitting antennas, limiting the maximum number of coherent sources to four. We conducted four sets of experiments for $n=1,2,3,$ and $4$ sources. In each set, 50 experimental results were measured by varying the angular interval of transmitters and the rotation stage angle. As shown in Fig.~\ref{fig_3}F(left), the AE-DNN achieved confidence values of $97.5\%$ in the source number estimation task. The simulation results demonstrated that the AE-DNN can successfully estimate the number of sources for up to 10 signal sources from different angles Fig.~\ref{fig_3}F(right). Evaluated on a comprehensive dataset of 10,000 simulated samples with varying source numbers and angular permutations, the AE-DNN attained an average confidence level of $98.3\%$ in source number estimation.

\subsection*{AE-DNN for integrated sensing, communication and computing}

In the study of 6G integrated sensing, communication and computing (ISCC), a key challenge lies in leveraging limited hardware and resources to collaboratively perform environmental sensing, communication transceiving, and multi-source separation in spatially dense multi-user cooperative environments \cite{zhuPushingAIWireless2023}. Based on the three-layer AE-DNN in Fig.~\ref{fig_3}A, we demonstrate ISCC high-capacity communication in the mmWave band with low latency, low energy, and high parallelism. AE-DNN performs sensing tasks by estimating the number and angles of signal sources, demixes free-space aliased BPSK and QAM16 signals to achieve light-speed sensing, and finally enables high-capacity communication through spatial multiplexing and beam focusing.

We first demonstrate that the beam focusing gain and broadband characteristics of AE-DNN dramatically enhance communication capacity. The AWG shown in Fig.~\ref{fig_3}B generates a QAM16 modulated signal (carrier frequency: 39 GHz, symbol rate: 100 MHz, sampling rate: 128 GSa/s). Due to far-field propagation loss, the signal received by the spectrum analyzer exhibits poor quality, as evidenced by the eye diagram in Fig.~\ref{fig_3}G(first column, top). By employing beam focusing, AE-DNN improves the SNR of the QAM16 signal, reducing the error vector magnitude (EVM) from –12.1 dB to –23.53 dB (Fig.~\ref{fig_3}G, first column). Further experiments measured the EVM curve of AE-DNN at various antenna angles ranging from –13.5$^\circ$ to 13.5$^\circ$ (Fig.~\ref{fig_3}H). The results show that without AE-DNN, the EVM remains at –12.4 dB; when AE-DNN is active, the EVM immediately drops to –22.4 dB, demonstrating an average improvement of 10 dB, substantially enhancing the SNR. To quantify the improvement in communication capacity, we measured the EVM as a function of symbol rate. The results demonstrate that at a given EVM level, i.e., under equivalent communication quality, the beam focusing gain and broadband characteristics of AE-DNN enable the communication system to support higher symbol rates. Statistical analysis indicates that within the EVM range of –23.5 dB to –13.5 dB, AE-DNN enables a 13.5-fold average increase in communication capacity.

At the same time, AE-DNN can achieve light-speed sensing and computing. The AWG simultaneously generates both QAM16 and BPSK signals at a center frequency of 39 GHz and a symbol rate of 100 MHz. Multiple sets of two-source experiments were conducted by adjusting the angular interval of the transmitting antennas and rotating the angular stage. Experimental results show that the confidence value of source number estimation and super-resolution DOA estimation reaches $96.0\%$ and $98.0\%$, respectively (Fig.~S5). When the angular interval is 3$^\circ$ (below the diffraction limit), conventional communication systems struggle to demodulate the aliased signals in free space, resulting in an EVM as high as –0.06 dB and -2.3 dB with QAM16 and BPSK demodulation, respectively (Fig.~\ref{fig_3}G, second column). Using the source separation capability of AE-DNN, aliased BPSK and QAM16 signals can be demixed to achieve light-speed computing. As shown in Fig.~\ref{fig_3}G(third column), the demixed QAM16 and BPSK signals exhibit clear eye diagrams with EVMs of –15.5 dB and –17.5 dB, respectively. When the angular interval is increased to 6$^\circ$, the EVMs of the demixed QAM16 and BPSK signals further improve to –21 dB and –20.5 dB (Fig.~\ref{fig_3}G, fourth column). By orthogonalizing the array response vectors in the analog domain, AE-DNN creates multiple parallel sub-channels for spatial multiplexing. Combined with beam focusing to enhance SNR, it ultimately achieves high-capacity communication, advancing the ISCC in 6G.

\begin{figure}[t!]
	\centering
	\includegraphics[width=1.0\textwidth]{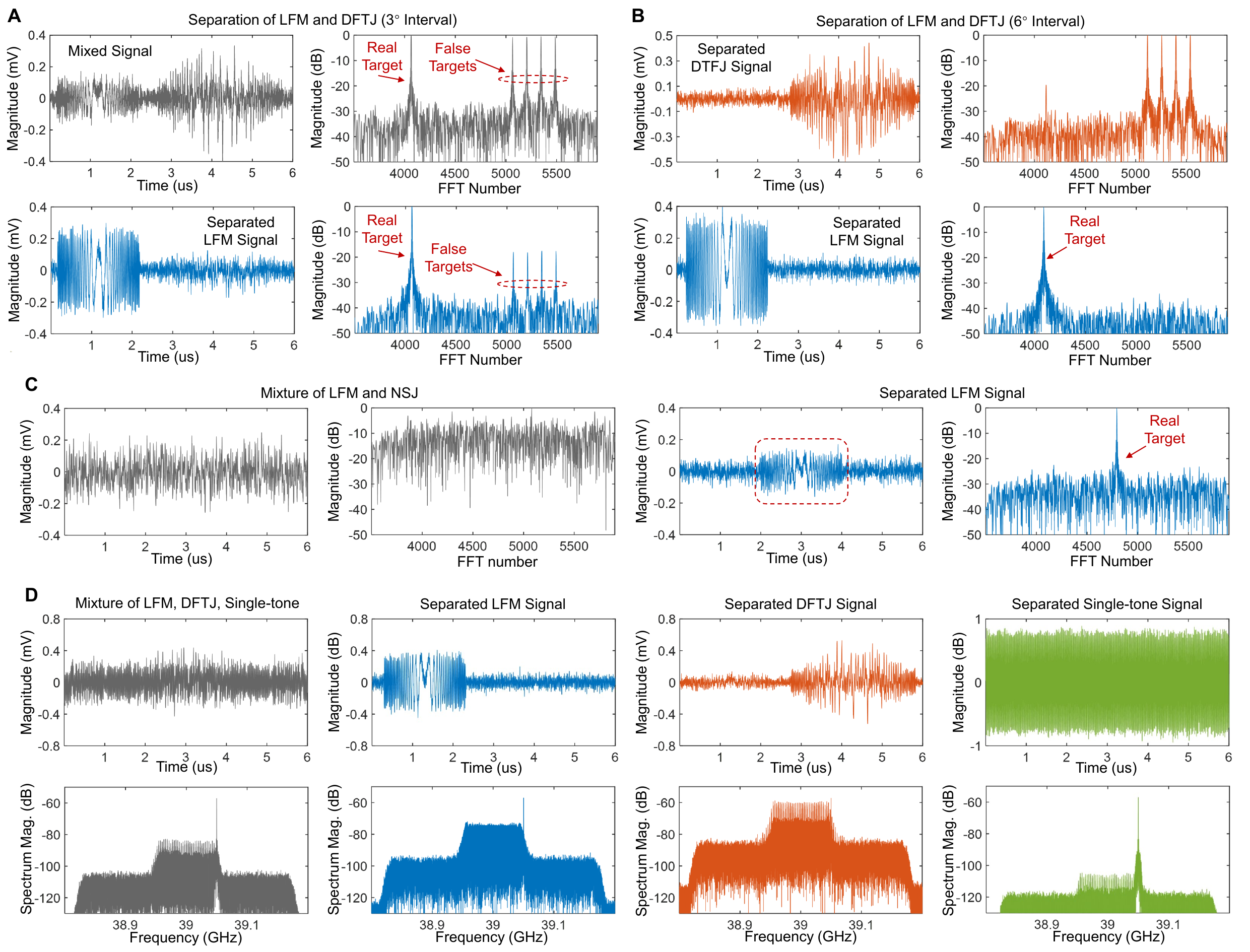}
	\caption{\textbf{Radar Anti-Jamming with Multiple Separation Channels.} (\textbf{A}) Time domain waveform and pulse compression results of the mixed LFM signal and the DFTJ (top). After AE-DNN source separation, the SNR of the LFM signal is improved, and the DFTJ is effectively suppressed (bottom). (\textbf{B}) When the angular interval is increased from 3° to 6°, the jamming suppression ratio of the demixed LFM channel reaches 35~dB. (\textbf{C}) Before separation, the LFM signal is overwhelmed by the high-power NSJ (left); after separation, the LFM signal waveform and pulse compression spikes are detectable (right). (\textbf{D}) Waveform and spectrum of the mixture of the LFM, DFTJ, and single-tone signal when three signals are simultaneously transmitted (top). After multi-channel source separation, the LFM, DFTJ, and single-tone signal are successfully demixed in both the time and frequency domains (bottom).}
	\label{fig_4}
\end{figure}

\subsection*{AE-DNN for multi-channel radar anti-jamming}

Based on the AE-DNN, we have realized radar anti-active jamming applications with light speed, high parallelism and low power consumption. In radar electronic warfare, dense false target jamming (DFTJ), an active deception jamming generated by digital radio frequency memory jammers \cite{soumekhSARECCMUsingPhaseperturbed2005}, can create numerous false targets to deceive radar receivers. Noise suppressive jamming (NSJ), with its broad bandwidth and high power, drowns out target echoes, making it difficult for radars to detect signals amidst noise. Conventional radar anti-jamming typically involves two steps: first, estimating the number and angles of signals and jamming using parameter estimation methods, and then applying adaptive beamforming to achieve spatial separation of jamming and desired signals. AE-DNN enables parallel execution of these tasks with significantly enhanced performance.

During the experiment, the AWG, shown in Fig.~\ref{fig_3}B, was used to simultaneously generate radar echo signals and active jamming, which are modulated as a linear frequency modulation (LFM) signal and DFTJ, respectively. The center frequency of both the LFM and DFTJ was 39~GHz, with a bandwidth of 100~MHz and a sampling rate of 128~GSa/s. The LFM signal and DFTJ were radiated through two transmitting antennas, propagated in the far field, received by AE-DNN and processed at the speed of light. The output of AE-DNN was analyzed using the FSW spectrum analyzer. The energy of ten detection regions was measured to accomplish source number estimation and super-resolution DOA estimation. The angular interval between the two transmitting antennas was set to 3$^\circ$ and 6$^\circ$, respectively, and multiple sets of two-source experiments were measured by rotating the angular stage. Experimental results showed that the confidence values of source number estimation and super-resolution DOA estimation reached $94.2\%$ and $98.0\%$, respectively (Fig.~S6). AE-DNN allows for accurately estimating the number and angles of signals and jamming.

Building on AE-DNN’s highly parallel source separation capability, effective separation of mixed signals and jamming was further achieved. Before separation, the received antenna detected the mixed signal of LFM and DFTJ (with 3$^\circ$ angular interval). The time-domain waveform and pulse compression results are shown in Fig.~\ref{fig_4}A(top). The DFTJ resulted in multiple false target peaks after pulse compression, severely hindering radar target identification. After source separation via AE-DNN, the LFM signal was demixed to the corresponding receiving antenna. Compared with signals before separation, the beam-focused LFM signal exhibited substantially enhanced amplitude, and the DFTJ was effectively suppressed with a jamming suppression ratio (JSR) of 18 dB (Fig.~\ref{fig_4}A, bottom). Statistical results from multiple sets of two-source experiments indicated an average JSR of 14.6 dB (Fig.~S7). When the angular interval between the radar signal and jamming was increased to 6$^\circ$, AE-DNN’s source separation performance improved further. Fig.~\ref{fig_4}B shows the time-domain and pulse compression results from receiving antennas number $\mathrm{Rx}_5$ and $\mathrm{Rx}_3$ after separation. $\mathrm{Rx}_5$ received only the LFM signal, with the DFTJ successfully suppressed with an JSR up to 35 dB. Meanwhile, the DFTJ was demixed to $\mathrm{Rx}_3$, which can be used for subsequent jamming waveform identification. The average JSR from multiple sets of two-source measurements was 20 dB (Fig.~S7). In addition to deception jamming, AE-DNN also suppressed wideband NSJ with power 18 dB higher than that of the LFM signal (Fig.~\ref{fig_4}C). Before separation, the LFM signal was overwhelmed by strong NSJ, and the radar could not detect the target signal even after pulse compression. After separation, the time-domain and pulse compression peak of the LFM became distinct and detectable, with an average JSR as high as 23 dB. Thus, AE-DNN effectively suppress both active deception jamming and noise suppression jamming, enabling highly parallel radar anti-jamming.

To verify AE-DNN’s multi-channel processing capability, three transmitting antennas with 6$^\circ$ interval were used, emitting the LFM signal, DFTJ, and a single-tone signal at 39.05 GHz, respectively. The single-tone signal was generated by VNA with a power 20 dB higher than the LFM and DFTJ. In Fig.~\ref{fig_4}D, the time-domain waveforms of the LFM and DFTJ were overwhelmed by the strong single-tone signal, and the spectra of the three signals were completely overlapped and indistinguishable. After source separation via AE-DNN, the three signals were demixed at the speed of light. The time-domain and spectra of each separated signal are shown in Fig.~\ref{fig_4}D. The results demonstrate that the three signals were successfully separated in both time and frequency domains, achieving effective channel crosstalk suppression. Besides, simulation validation further shows that AE-DNN can support up to 10 source separation channels (supplementary text 2.3).

\begin{table}[htbp]
\centering
\caption{Comparison AE-DNN with state-of-the-art optoelectronic and electronic computing architectures}
\label{tab_1}
\renewcommand{\arraystretch}{1.2}
\scriptsize 
\begin{tabular}{|p{1.5cm}|p{1.1cm}|p{1.6cm}|p{1.0cm}|p{1.9cm}|p{1.0cm}|p{1.2cm}|p{0.9cm}|p{2.3cm}|}
\hline
\textbf{Processor Type} & \textbf{Total Neurons} & \textbf{Computing Paradigm} & \textbf{Aperture} & \textbf{Resolution} & \textbf{Channels} & \textbf{Nonlinear} & \textbf{Gain/ Loss} & \textbf{Sensing and Computing Tasks} \\
\hline
AE-DNN & 6075 & Fully-analog (1.67 µs) & 3D & Scalable \, \, \, \, super-resolution & 10 & Encoding nonlinear & +13 dB & Source number and \, \, DOA estimation, \, \, source separation. \\
\hline
MRR \, \, \, \, \, \, Array \cite{zhangSystemonchipMicrowavePhotonic2024} & 4 & Optoelectronic (33 ms) & 1D & Diffraction-limited & 2 & None & -15 dB & Blind source \, \, \, \, \, \, separation \\
\hline
MZI \, \, \, \, \, \, Array \cite{seyedinnavadehDeterminingOptimalCommunication2023} & 30 & Optoelectronic (50 ms) & 1D & Diffraction-limited & 2 & None & -2 dB & Source separation \\
\hline
Holographic\, \, \, RIS \cite{zhu2025selfcontrolleda} & 1024 & Optoelectronic (6.2 ms) & 2D & Diffraction-limited & 2 & None & -0.5 dB & DOA estimation \\
\hline
S-DNN \cite{gaoSuperresolutionDiffractiveNeural2024} & 4096 & Fully-analog & 2D & Super-resolution & 2 & None & N/A & DOA estimation \\
\hline
Classic \, \, \, MUSIC \cite{kgrampsrddf5gts} & N/A & Electronic \, ~~(1 ms) & 2D & Super-resolution & N/A & Electronic nonlinear & N/A & DOA estimation \\
\hline
\end{tabular}
\end{table}

\subsection*{Discussion}

We systematically compare AE-DNN with state-of-the-art sensing and computing methods based on optoelectronic fusion architectures and the latest electronic commercial product, as summarized in Table 1. The results demonstrate that AE-DNN achieves a total latency of merely 1.67 microseconds for parallel execution of three core sensing tasks, representing an improvement of approximately three orders of magnitude over its electronic counterpart. Specifically, the VNA has the fastest detection speed of 67 ns, and the RF switch, e.g., TLSP10T26.5G40GA (Talent Microwave Inc.) achieves a switching speed of 100 ns for ten detection regions, resulting in a total response speed of 1.67 microseconds. While classical MUSIC can only handle incoherent signals, AE-DNN directly processes EM fields in the analog domain to establish the independent channel mapping between sources and detectors, enabling it to handle multiple coherent or incoherent sources. Furthermore, AE-DNN operates with negligible power consumption during analog computation, constituting a fundamentally passive sensing system. Integrating a total of 6075 neurons, AE-DNN can process multiple aliased signals sharing identical temporal, frequency, and polarization characteristics. It currently supports up to 10 separation channels, with the potential for further scaling to enhance channel number. Notably, leveraging its beam-focusing effect, AE-DNN provides a SNR gain of approximately 13 dB, whereas other optoelectronic fusion schemes exhibit varying degrees of signal loss. In addition, the pioneering 3D aperture engineering and encoding nonlinearity of AE-DNN enable scalable super-resolution perception and sophisticated computing tasks.

While AE-DNN, empowered with encoding and square-law detection nonlinearities, achieves remarkable efficiency and operational speed in wireless signal processing, the inherent linearity of wave superposition constrains their computational expressivity for complex tasks. The incorporation of nonlinear metasurfaces \cite{luo2019intensitydependent, kiani2020spatial, ning2024reprogrammable, ning2025multilayer} for multidimensional EM field activation, e.g., gating and thresholding, therefore represents a compelling and increasingly mature route toward enhancing the functional intelligence of such physical computing architectures. In particular, self-biased passive designs enable intensity-dependent wave manipulation without external power supplies, offering a practical means of introducing nonlinearity while preserving system compactness and energy efficiency \cite{kiani2020spatial}. Moreover, the integration of radio-frequency power amplifiers and detection circuits allows hardware-level ReLU activation functions to be implemented, providing nanosecond-scale response times together with programmable activation thresholds \cite{ning2025multilayer}. Embedding these nonlinear mechanisms into existing diffractive computing frameworks is expected to substantially expand their representational capacity for boosting task performance, thereby enabling next-generation autonomous EM platforms capable of adaptive signal modulation and universal computing at the physical layer.

The scalability of the AE-DNN architecture allows for a dynamic expansion of the effective aperture $D_{\mathrm{eff}}$, which fundamentally redefines the sensing boundaries for diverse environments. In practical scenarios, where the receiver is positioned hundreds of meters to several kilometers away from the transmitter, the system often operates in the regime where the distance $R {>} 2D_{\mathrm{eff}}^2/\lambda_c$. Under this condition, the incident wave at the receiver satisfies the criteria of a large-aperture far-field plane wave. In this far-field regime, the primary advantage of AE-DNN’s expanded aperture is the significant enhancement of angular resolution for the superior spatial separation of closely spaced sources by recognizing the incident EM field patterns, even at kilometer-scale ranges. However, the quadratic dependence of the Fraunhofer distance on the effective aperture ($D_{\mathrm{eff}}^2$) implies that as AE-DNN scales up its aperture with 3D aperture engineering framework to capture more energy and improve resolution, the far-field boundary is pushed further. For objects or targets that fall within this expanded boundary ($R {<} 2D_{\mathrm{eff}}^2/\lambda_c$), the plane-wave assumption no longer holds. In these near-field regions, AE-DNN can shift its processing paradigm to leverage wavefront curvature. Unlike traditional systems that may suffer from model mismatch in the near-field, AE-DNN’s parallel execution capabilities with MSA training design method allow it to process spherical wave models in real-time. This facilitates accurate depth estimation and 3D localization by extracting information from the non-linear encoded phase distributions across the large aperture. Therefore, this research direction could evolve AE-DNN into a robust and unified sensing framework that maintains high-fidelity plane-wave processing for long-range surveillance while simultaneously enabling precision range-dependent perception for proximal targets.

In summary, our proposed 3D aperture engineering framework can capture obliquely incident plane waves from a broader spatial ranges, achieving an angular resolution ${\sim}N$ times greater than the single-aperture Rayleigh diffraction limit through $N$-layer deeply cascaded metasurfaces. AE-DNN utilizes MSA training and encoding nonlinearity to orthogonalize the array response vector in the analog domain, thus enabling fully analog sensing and computing. In the 36-41 GHz range, AE-DNN can process multiple coherent signal sources in parallel, achieving super-resolution DOA estimation, source number estimation, and source separation with channel isolation exceeding 20 dB. In ISCC high-capacity communication, AE-DNN separates overlapped BPSK and QAM-16 signals for spatial multiplexing, increasing channel capacity by an average factor of 13.5 via beam focusing and broadband operation. In radar anti-jamming application, AE-DNN perceives the number and angles of radar signals and jamming at light speed, enhancing the signal and effectively suppressing deception and suppressive jamming through source separation. Furthermore, with up to 10 separation channels, AE-DNN's scalability far exceeds that of on-chip solutions. Crucially, the proposed 3D aperture engineering reveals a universal paradigm: any physical sensor based on wave phenomena (e.g., EM or acoustic waves) can surpass the perception limit of a single-layer physical aperture through such designs, and achieve significantly enhanced and scalable sensing resolution.




\clearpage 

%

\begin{thebibliography}{10}
\providecommand{\url}[1]{\texttt{#1}}
\expandafter\ifx\csname urlstyle\endcsname\relax
  \providecommand{\doi}[1]{doi:\discretionary{}{}{}#1}\else
  \providecommand{\doi}{doi:\discretionary{}{}{}\begingroup \urlstyle{rm}\Url}\fi

\bibitem{javadiRadarNetworksReview2020}
S.~H. Javadi, A.~Farina, Radar Networks: A Review of Features and Challenges. \emph{Information Fusion} \textbf{61}, 48--55 (2020).

\bibitem{dangWhatShould6G2020}
S.~Dang, O.~Amin, B.~Shihada, M.-S. Alouini, What Should 6G Be? \emph{Nature Electronics} \textbf{3}~(1), 20--29 (2020).

\bibitem{liuEdgeComputingAutonomous2019}
S.~Liu, \emph{et~al.}, Edge Computing for Autonomous Driving: Opportunities and Challenges. \emph{Proceedings of the IEEE} \textbf{107}~(8), 1697--1716 (2019).

\bibitem{9149671}
C.~Li, P.~Qi, D.~Wang, Z.~Li, On the Anti-Interference Tolerance of Cognitive Frequency Hopping Communication Systems. \emph{IEEE Transactions on Reliability} \textbf{69}~(4), 1453--1464 (2020).

\bibitem{9394593}
D.~Mao, \emph{et~al.}, An Efficient Anti-Interference Imaging Technology for Marine Radar. \emph{IEEE Transactions on Geoscience and Remote Sensing} \textbf{60}, 1--13 (2022).

\bibitem{krimTwoDecadesArray1996a}
H.~Krim, M.~Viberg, Two Decades of Array Signal Processing Research: The Parametric Approach. \emph{IEEE Signal Processing Magazine} \textbf{13}~(4), 67--94 (1996).

\bibitem{munozEnhancingFibreopticDistributed2022}
F.~Mu{\~n}oz, M.~A. Soto, Enhancing Fibre-Optic Distributed Acoustic Sensing Capabilities with Blind near-Field Array Signal Processing. \emph{Nature Communications} \textbf{13}~(1), 4019 (2022).

\bibitem{wangOverviewEnhancedMassive2019}
M.~Wang, F.~Gao, S.~Jin, H.~Lin, An Overview of Enhanced Massive MIMO with Array Signal Processing Techniques. \emph{IEEE Journal of Selected Topics in Signal Processing} \textbf{13}~(5), 886--901 (2019).

\bibitem{devosLOFARTelescopeSystem2009}
M.~{de Vos}, A.~W. Gunst, R.~Nijboer, The LOFAR Telescope: System Architecture and Signal Processing. \emph{Proceedings of the IEEE} \textbf{97}~(8), 1431--1437 (2009).

\bibitem{wangOnchipTopologicalBeamformer2024}
W.~Wang, \emph{et~al.}, On-Chip Topological Beamformer for Multi-Link Terahertz 6G to XG Wireless. \emph{Nature} \textbf{632}~(8025), 522--527 (2024).

\bibitem{tanDirectionArrivalEstimation2014}
Z.~Tan, Y.~C. Eldar, A.~Nehorai, Direction of Arrival Estimation Using Co-Prime Arrays: A Super Resolution Viewpoint. \emph{IEEE Transactions on Signal Processing} \textbf{62}~(21), 5565--5576 (2014).

\bibitem{moreiraTutorialSyntheticAperture2013}
A.~Moreira, \emph{et~al.}, A Tutorial on Synthetic Aperture Radar. \emph{IEEE Geoscience and Remote Sensing Magazine} \textbf{1}~(1), 6--43 (2013).

\bibitem{linArtificialIntelligenceBuilt2022}
X.~Lin, Artificial Intelligence Built on Wireless Signals. \emph{Nature Electronics} \textbf{5}~(2), 69--70 (2022).

\bibitem{wangParallelInmemoryWireless2023}
C.~Wang, \emph{et~al.}, Parallel In-Memory Wireless Computing. \emph{Nature Electronics} \textbf{6}~(5), 381--389 (2023).

\bibitem{huangRadiofrequencySignalProcessing2025}
Y.~Huang, \emph{et~al.}, Radiofrequency Signal Processing with a Memristive System-on-a-Chip. \emph{Nature Electronics} pp. 1--10 (2025).

\bibitem{wangSimplifiedRadarArchitecture2025}
S.~R. Wang, \emph{et~al.}, Simplified Radar Architecture Based on Information Metasurface. \emph{Nature Communications} \textbf{16}~(1), 6505 (2025).

\bibitem{wetzsteinInferenceArtificialIntelligence2020a}
G.~Wetzstein, \emph{et~al.}, Inference in Artificial Intelligence with Deep Optics and Photonics. \emph{Nature} \textbf{588}~(7836), 39--47 (2020).

\bibitem{shastriPhotonicsArtificialIntelligence2021}
B.~J. Shastri, \emph{et~al.}, Photonics for Artificial Intelligence and Neuromorphic Computing. \emph{Nature Photonics} \textbf{15}~(2), 102--114 (2021).

\bibitem{shenDeepLearningCoherent2017}
Y.~Shen, \emph{et~al.}, Deep Learning with Coherent Nanophotonic Circuits. \emph{Nature Photonics} \textbf{11}~(7), 441--446 (2017).

\bibitem{xu202111}
X.~Xu, \emph{et~al.}, 11 TOPS Photonic Convolutional Accelerator for Optical Neural Networks. \emph{Nature} \textbf{589}~(7840), 44--51 (2021).

\bibitem{feldmann2021parallel}
J.~Feldmann, \emph{et~al.}, Parallel Convolutional Processing Using an Integrated Photonic Tensor Core. \emph{Nature} \textbf{589}~(7840), 52--58 (2021).

\bibitem{pai2023experimentally}
S.~Pai, \emph{et~al.}, Experimentally Realized in Situ Backpropagation for Deep Learning in Photonic Neural Networks. \emph{Science} \textbf{380}~(6643), 398--404 (2023).

\bibitem{sludds2022delocalized}
A.~Sludds, \emph{et~al.}, Delocalized Photonic Deep Learning on the Internet's Edge. \emph{Science} \textbf{378}~(6617), 270--276 (2022).

\bibitem{bogaertsProgrammablePhotonicCircuits2020}
W.~Bogaerts, \emph{et~al.}, Programmable Photonic Circuits. \emph{Nature} \textbf{586}~(7828), 207--216 (2020).

\bibitem{fengIntegratedLithiumNiobate2024}
H.~Feng, \emph{et~al.}, Integrated Lithium Niobate Microwave Photonic Processing Engine. \emph{Nature} \textbf{627}~(8002), 80--87 (2024).

\bibitem{xuLargescalePhotonicChiplet2024}
Z.~Xu, \emph{et~al.}, Large-Scale Photonic Chiplet Taichi Empowers 160-TOPS/W Artificial General Intelligence. \emph{Science} \textbf{384}~(6692), 202--209 (2024).

\bibitem{dongPartialCoherenceEnhances2024}
B.~Dong, \emph{et~al.}, Partial Coherence Enhances Parallelized Photonic Computing. \emph{Nature} \textbf{632}~(8023), 55--62 (2024).

\bibitem{milanizadehSeparatingArbitraryFreespace2022}
M.~Milanizadeh, \emph{et~al.}, Separating Arbitrary Free-Space Beams with an Integrated Photonic Processor. \emph{Light: Science \& Applications} \textbf{11}~(1), 197 (2022).

\bibitem{zhangSystemonchipMicrowavePhotonic2024}
W.~Zhang, \emph{et~al.}, A System-on-Chip Microwave Photonic Processor Solves Dynamic RF Interference in Real Time with Picosecond Latency. \emph{Light: Science \& Applications} \textbf{13}~(1), 14 (2024).

\bibitem{seyedinnavadehDeterminingOptimalCommunication2023}
S.~SeyedinNavadeh, \emph{et~al.}, Determining the Optimal Communication Channels of Arbitrary Optical Systems Using Integrated Photonic Processors. \emph{Nature Photonics} pp. 1--7 (2023).

\bibitem{linAllopticalMachineLearning2018a}
X.~Lin, \emph{et~al.}, All-Optical Machine Learning Using Diffractive Deep Neural Networks. \emph{Science} \textbf{361}~(6406), 1004--1008 (2018).

\bibitem{zhouLargescaleNeuromorphicOptoelectronic2021}
T.~Zhou, \emph{et~al.}, Large-Scale Neuromorphic Optoelectronic Computing with a Reconfigurable Diffractive Processing Unit. \emph{Nature Photonics} \textbf{15}~(5), 367--373 (2021).

\bibitem{chenAllanalogPhotoelectronicChip2023}
Y.~Chen, \emph{et~al.}, All-Analog Photoelectronic Chip for High-Speed Vision Tasks. \emph{Nature} \textbf{623}~(7985), 48--57 (2023).

\bibitem{gaoSuperresolutionDiffractiveNeural2024}
S.~Gao, \emph{et~al.}, Super-Resolution Diffractive Neural Network for All-Optical Direction of Arrival Estimation beyond Diffraction Limits. \emph{Light: Science \& Applications} \textbf{13}~(1), 161 (2024).

\bibitem{yuAllopticalImageTransportation2025}
H.~Yu, \emph{et~al.}, All-Optical Image Transportation through a Multimode Fibre Using a Miniaturized Diffractive Neural Network on the Distal Facet. \emph{Nature Photonics} pp. 1--8 (2025).

\bibitem{yu2011light}
N.~Yu, \emph{et~al.}, Light Propagation with Phase Discontinuities: Generalized Laws of Reflection and Refraction. \emph{Science} \textbf{334}~(6054), 333--337 (2011).

\bibitem{qianDynamicRecognitionMirage2022}
C.~Qian, \emph{et~al.}, Dynamic Recognition and Mirage Using Neuro-Metamaterials. \emph{Nature Communications} \textbf{13}~(1), 2694 (2022).

\bibitem{liuProgrammableDiffractiveDeep2022}
C.~Liu, \emph{et~al.}, A Programmable Diffractive Deep Neural Network Based on a Digital-Coding Metasurface Array. \emph{Nature Electronics} \textbf{5}~(2), 113--122 (2022).

\bibitem{zhu2025selfcontrolleda}
J.~Zhu, Z.~Gu, Q.~Ma, L.~Dai, T.~J. Cui, A Self-Controlled Reconfigurable Intelligent Surface Inspired by Optical Holography. \emph{Nature Electronics} \textbf{8}~(11), 1108--1118 (2025).

\bibitem{woodPhysicalOpticsRobert1919}
R.~W. Wood, {Burndy Library}, \emph{Physical Optics: By Robert W. Wood} (New York) (1919).

\bibitem{zhuPushingAIWireless2023}
G.~Zhu, \emph{et~al.}, Pushing AI to Wireless Network Edge: An Overview on Integrated Sensing, Communication, and Computation towards 6G. \emph{Science China Information Sciences} \textbf{66}~(3), 130301 (2023).

\bibitem{soumekhSARECCMUsingPhaseperturbed2005}
M.~Soumekh, SAR-ECCM Using Phase-Perturbed LFM Chirp Signals and DRFM Repeat Jammer Penalization, in \emph{IEEE International Radar Conference, 2005.} (2005), pp. 507--512.

\bibitem{kgrampsrddf5gts}
R.~Schwarz, DDF5GTS High Speed Scanning Direction Finder, https://www.rohde-schwarz.com/us/products/aerospace-defense-security/rack-mount-multi-channel/rs-ddf5gts-high-speed-scanning-direction-finder\_63493-78912.html.

\bibitem{luo2019intensitydependent}
Z.~Luo, \emph{et~al.}, Intensity-Dependent Metasurface with Digitally Reconfigurable Distribution of Nonlinearity. \emph{Advanced Optical Materials} \textbf{7}~(19), 1900792 (2019).

\bibitem{kiani2020spatial}
M.~Kiani, A.~Momeni, M.~Tayarani, C.~Ding, Spatial Wave Control Using a Self-Biased Nonlinear Metasurface at Microwave Frequencies. \emph{Optics Express} \textbf{28}~(23), 35128--35142 (2020).

\bibitem{ning2024reprogrammable}
Y.~M. Ning, Q.~Ma, Q.~Xiao, Z.~Gu, T.~J. Cui, Reprogrammable Nonlinear Transmission Controls Using an Information Metasurface. \emph{Advanced Optical Materials} \textbf{12}~(3), 2301525 (2024).

\bibitem{ning2025multilayer}
Y.~M. Ning, \emph{et~al.}, Multilayer Nonlinear Diffraction Neural Networks with Programmable and Fast ReLU Activation Function. \emph{Nature Communications} \textbf{16}~(1), 10332 (2025).

\end{thebibliography}
\bibliographystyle{sciencemag}

%
%


\section*{Acknowledgments}

\paragraph*{Funding:}
This work is supported by the National Key Research and Development Program of China (No. 2021ZD0109902) and the National Natural Science Foundation of China (No. 62275139 and No. 62525112).

\paragraph*{Author contributions:}
X.L. and Y.S. initiated and supervised the project. X.L. and Y.S. supported resources and funding. X.L., S.G., and Y.S. conceived the research and designed the methods. S.G. implemented the AE-DNN. S.G., H.Z., and X.L. constructed the experimental setups. S.G., S.Y., and X.L. implemented the algorithm and conducted experiments. S.G., S.Y., and H.Z. processed the data. X.L., S.G., S.Y., and Y.S. analyzed and interpreted the results. S.G., X.L., and Y.S. wrote the manuscript with input from all authors. All authors contributed to the discussion.

\paragraph*{Competing interests:}
There are no competing interests to declare.

\paragraph*{Data and materials availability:}
All data needed to evaluate the conclusions are present in the main text or supplementary materials. The data repository is available on GitHub via the following link: https://github.com/THPCILab/AE-DNN

\paragraph*{Code availability:}
The code for MSA training of 5-layer AE-DNN are available on GitHub via the following link: https://github.com/THPCILab/AE-DNN




\newpage

\renewcommand{\thefigure}{S\arabic{figure}}
\renewcommand{\thetable}{S\arabic{table}}
\renewcommand{\theequation}{S\arabic{equation}}
\renewcommand{\thepage}{S\arabic{page}}
\setcounter{figure}{0}
\setcounter{table}{0}
\setcounter{equation}{0}
\setcounter{page}{1} 


\begin{center}
\section*{Supplementary Materials for\\ 3D aperture-engineered diffractive neural networks \\ for super-resolution electromagnetic wave computing}


\normalsize Sheng Gao$^{1}$, Songtao Yang$^{1}$, Haiou Zhang$^{1}$, Yuan Shen$^{1,2,\ast}$, and Xing Lin$^{1,2,\ast}$\\
\normalsize $^{1}$Department of Electronic Engineering, Tsinghua University, Beijing, 100084, China.\\
\normalsize $^{2}$Beijing National Research Center for Information Science and Technology, \\
\normalsize Tsinghua University, Beijing, 100084, China.\\
\normalsize $^\ast$Corresponding authors. Email: lin-x@tsinghua.edu.cn, shenyuan$\_$ee@tsinghua.edu.cn\\

\end{center}

\subsubsection*{This PDF file includes:}
Materials and Methods\\
Supplementary text\\
Figures S1 to S9\\
Tables S1 to S2\\



\clearpage
\newpage


\subsection*{1 \quad Materials and Methods}

\subsubsection*{1.1 \quad Mini-FCNN structures}

To ensure accurate and differentiable prediction of the broadband amplitude and phase responses of metasurface units, the AE-DNN framework employs miniature fully connected neural networks (mini-FCNNs). As illustrated in Fig.~S1A, the mini-FCNN is designed to map the geometric parameters to the high-dimensional electromagnetic response space. The input is defined as a two-dimensional geometric parameter vector $\mathbf{x} = (d_a, d_b)^{T}$. This vector is first projected into a high-dimensional feature space using a learnable projection matrix $\mathbf{B} \in \mathbb{R}^{128 \times 2}$. The projection is computed as $\mathbf{z} = 2\pi \mathbf{B}\mathbf{x}$, and the Fourier embedding is subsequently constructed as $\gamma(\mathbf{x}) = [\sin(\mathbf{z});\cos(\mathbf{z})]\in \mathbb{R}^{256 \times 1}$. This 256-dimensional embedded feature vector $\gamma(\mathbf{x})$ is then processed through a fully connected hidden layer with ReLU activation. The final output layer produces a 1003-dimensional vector, which represents the broadband amplitude or phase responses sampled from 25~GHz to 55~GHz with a spectral resolution of 0.03~GHz.

The training of the mini-FCNN utilizes a comprehensive response library comprising 10,201 distinct unit cell configurations. This dataset defines a uniform grid over the geometric parameters $d_a$ and $d_b$, spanning from 1.4~mm to 2.4~mm with a fine resolution of 0.01~mm, where the corresponding ground-truth electromagnetic responses, i.e., amplitude and phase modulations of polarized EM field across broadband frequency ranges, were obtained via full-wave simulations. During training, the mean squared error (MSE) between the predicted responses and ground truth was used as the loss function, and the network was optimized using the Adam optimizer. Additionally, to prevent the network from overfitting, L2 regularization with a weight decay coefficient of $1\times10^{-6}$ was applied. As shown in Fig.~S1B, the comparison at 39~GHz demonstrates that the mini-FCNN achieves excellent agreement with the full-wave simulation results. The network achieves high-fidelity reconstruction of complex amplitude and phase distributions spanning the entire geometric parameter space, validating its reliability as a surrogate solver for gradient-based optimization tasks. Fig. S1C shows the mean absolute error of the amplitude and phase predictions of mini-FCNN as a function of frequency. The results validate that mini-FCNN can accurately predict the modulation response of metasurface units across broadband frequency ranges.

\subsubsection*{1.2 \quad Fabrication of metasurfaces}

The fabrication of metasurfaces comprising meta-atom arrays was achieved using high-precision standard printed circuit board (PCB) fabrication techniques. This manufacturing approach was utilized due to its reliability and scalability with a feature size tolerance of $0.025 \, \text{mm}$. In our metasurface design, the geometric parameters $d_a$ and $d_b$ are varied across a total design space of $1.0 \, \text{mm}$ to facilitate the required phase modulation range. Our design achieves a full 0$\sim$2$\pi$ phase coverage, allowing for complete control over the wavefront. The precision of the resulting optical phase control is determined by the ratio of the fabrication tolerance to the total tunable range. With a tolerance-to-range ratio of $1:40$, the meta-atom supports a phase modulation bit depth exceeding 5 bits ($2^5 = 32$ levels). This high degree of modulation discretization at each layer is sufficient to adapt to the model quantization and maintain the performance of sensing and computing tasks.

\subsubsection*{1.3 \quad Experimental systems}

To rigorously evaluate the broadband performance and multi-functional signal-processing capabilities of the proposed AE-DNN, a sophisticated experimental platform was established within a microwave anechoic chamber (Fig.~S9). The AE-DNN was securely mounted on a high-precision angular rotation stage, with its rotation axis precisely aligned at the center of the metasurface. Driven by a dedicated stepper motor, the stage provided a full $360^\circ$ azimuthal range with a rotation accuracy of $0.01^\circ$, facilitating meticulous control over the incident plane-wave angles.

The signal generation subsystem utilized a dual-channel Keysight M8198 Arbitrary Waveform Generator (AWG) to synthesize independent, high-fidelity source signals. This allowed for the simultaneous transmission of complex radar waveforms and communication sequences (such as QPSK or 16-QAM) with varying modulation formats. These signals were connected to multiple far-field horn antennas, ensuring uniform plane-wave illumination across the AE-DNN aperture.

For the characterization of the output energy distribution, a waveguide probe was integrated onto a high-resolution XY mechanical platform. This platform was driven by two vertically positioned stepper motors, enabling precise movement in both horizontal and vertical directions over a range of $65\text{ cm}$ with a positioning accuracy of $0.01\text{ mm}$. A systematic scanning procedure was executed across ten designated detection regions with a fine step size of $\lambda_0/8$. The probe was connected to an R\&S FSW spectrum analyzer for real-time demodulation and spectral analysis, as well as a Vector Network Analyzer (VNA) for fundamental electromagnetic field measurements.

The entire system was orchestrated by a customized scanning program that synchronized the communication between the angular rotation stage, the XY translation platform, and the VNA. This automation enabled the precise acquisition of output energy distributions corresponding to different incident angles and signal configurations.

\clearpage
\newpage

\subsection*{2 \quad Supplementary text}

\subsubsection*{2.1 \quad AE-DNN performance under varying network settings}

Leveraging the 3D aperture engineering framework and nonlinear sensing mechanism, AE-DNN can progressively enhance its angular resolution—achieving scalable super-resolution sensing—by increasing the number of cascaded layers to expand the effective receiving aperture. To validate this capability, we conducted systematic simulations with the following configuration: each metasurface layer contained a fixed array of 23$\times$23 unit cells with a unit size of $3\,mm$, corresponding to a physical aperture size of $D = 8.97\lambda_c$ by setting central frequency $f_c = 39\,\text{GHz}$. The inter-layer spacing was set to $L = 9\lambda_c$. Far-field sources were incident at angles greater than 45$^\circ$. Under this setup, the equivalent aperture expansion achieved by AE-DNN through wavefront propagation is given by $\Delta D = (N-1)L \cdot \tan{\theta}$. The angular resolution was determined using a criterion that the inter-channel crosstalk suppression ratio must exceed 7 dB, ensuring a DOA estimation accuracy greater than $95\%$ for any two sources within the specified angular range.

Figs.~2D and S2 illustrate the performance of AE-DNN's angular resolution with an increasing number of layers. It is evident that as the layer count rises from 1 to 10, the angular resolution improves continuously from 6.4$^\circ$ to 0.6$^\circ$, demonstrating clear scalable super-resolution sensing. Notably, the 10-layer AE-DNN achieves an angular resolution that surpasses the diffraction limit of a single-layer aperture by a factor of $\sim$10. In contrast, the conventional S-DNN, limited by its single-layer two-dimensional aperture, shows negligible resolution improvement with added layers. Furthermore, because the effective input aperture of AE-DNN significantly exceeds the base size of $8.97\lambda$, the energy of its focused beam does not suffer significant attenuation with increasing layer number and propagation distance, thereby maintaining high beam focusing gain.

Fig.~S2 further present the variation of AE-DNN's angular resolution with the aperture size (i.e., the number of unit cells) per layer within a five-layer architecture. Simulation results indicate that as the single-layer aperture expands and its corresponding diffraction-limited resolution decreases, AE-DNN—by virtue of its 3D aperture engineering framework—consistently delivers super-resolution performance approximately five times better than the corresponding diffraction limit across various aperture sizes, while simultaneously maintaining high energy efficiency.

\subsubsection*{2.2 \quad Numerical evaluations of three-layer AE-DNN}
In addition to the angular spectrum method (ASM) for numerical modeling, we employed CST Studio Suite to evaluate the performance of the \text{AE-DNN} for more accurate simulations. The commercial software CST Microwave Studio enables precise three-dimensional full-wave EM simulations based on the finite integration method, thereby ensuring the excellent experimental performance of the \text{AE-DNN}. The trained \text{AE-DNN} comprises three cascaded metasurfaces composed of $45 \times 45$ meta-atoms. Each meta-atom consists of multiple identical copper structures and dielectric layers. As clarified in supplementary text~1.1, the intrinsic electromagnetic response of each meta-atom is  determined by its geometric parameters $\left(d_a,\, d_b\right)$ under a fixed orientation, which govern the co-polarized amplitude and phase response. To fully exploit the anisotropic nature of the meta-atom, an additional rotational degree of freedom $\alpha$ is introduced. Through a standard coordinate transformation of the corresponding Jones matrix, the physical rotation of the unit cell enables continuous and independent modulation of the effective amplitude and phase experienced by the incident electromagnetic wave. Specifically, this modulation can be formalized as:
\begin{equation}
\mathbf{E}_{out} = \mathbf{M}(d_a, d_b, \alpha) \mathbf{E}_{inc} = \mathbf{R}(\alpha) \mathbf{J}_{int}(d_a, d_b) \mathbf{R}^{-1}(\alpha) \mathbf{E}_{inc}
\end{equation}
where $\mathbf{J}_{int}(d_a, d_b) = \text{diag}(\tilde{t}_u, \tilde{t}_v)$ represents the intrinsic response along the principal axes (with $\tilde{t}_{u,v}$ being the complex transmission coefficients), and $\mathbf{R}(\alpha)$ denotes the rotation matrix:

\begin{equation}
\mathbf{R}(\alpha) = \begin{pmatrix} \cos\alpha & -\sin\alpha \\ \sin\alpha & \cos\alpha \end{pmatrix}
\end{equation}
Accordingly, the combined parameter set $\left(d_a,\, d_b,\, \alpha\right)$ establishes a complete and physically informed mapping from the meta-atom geometry to its complex electromagnetic modulation, with the corresponding geometric parameters schematically illustrated in Fig.~S4A. Since the broadband robust training strategy directly optimizes these geometric parameters for each meta-atom, the trained \text{AE-DNN} can be straightforwardly modeled in CST using the resulting $45 \times 45 \times 3$ parameter sets. The constructed CST model and the corresponding unit-cell structures of the metasurfaces are shown in Fig.~S4B.


The CST simulations employed open-space boundary conditions. Plane waves with various incident angles were used as inputs, and the resulting field distributions of the \text{AE-DNN} were taken as outputs. The frequency range was set from $36$~GHz to $41$~GHz. Fig.~S4C compares the output field distributions obtained from the ASM and CST simulations under illumination by a plane wave at $-7.5^{\circ}$. The results show close agreement between the two methods. Furthermore, simulations were conducted with plane wave incident angles ranging from $-13.5^{\circ}$ to $13.5^{\circ}$ in steps of $3^{\circ}$. The energy distribution matrix obtained from CST simulations is presented in Fig.~S4D. Since CST does not support the simultaneous excitation by two plane waves, the complex amplitudes of the output field distributions from single-source simulations were coherently superimposed pairwise. This allowed us to reconstruct the energy distribution matrix for $45$ two-source angular permutations. The results (Fig.~S4F) demonstrate that the \text{AE-DNN} successfully achieves super-resolution DOA estimation with high confidence.

\subsubsection*{2.3 \quad AE-DNN supports up to 10 source separation channels}

Leveraging its array signal processing capability, \text{AE-DNN} enables flexible manipulation of beam direction, thereby routing EM waves incident from different angles to independent detection regions on the output plane in the wave domain. This allows for real-time parallel processing of multiple input signals without relying on waveguide routing structures for on-chip photonic computing. With $10$ detection regions integrated on the output plane, the current upper limit of separable channels in \text{AE-DNN} is $10$; this capacity can be further expanded by incorporating additional detection regions.

In our simulations, $10$ LFM signals with distinct incident angles and time delays were generated. The time-domain waveform and pulse compression results of their coherently superimposed mixed signal are shown in Fig.~S8A. It can be observed that the mixed LFM signals are indistinguishable, and pulse compression produces $10$ peaks of comparable magnitude. After light-speed source separation via \text{AE-DNN}, the time-domain waveforms and pulse compression outcomes of the demixed signals from each separation channel are presented in Fig.~S8B. Owing to the beam focusing gain of \text{AE-DNN}, the amplitude of the LFM signal within the corresponding angle interval of each separation channel is significantly improved, yielding clear waveforms. Each channel exhibits only one dominant peak after pulse compression, while the other nine LFM signals are effectively suppressed, achieving an average channel crosstalk suppression ratio of $17.6$~dB.

Compared to existing on-chip beam separation technologies (\emph{27, 28}), \text{AE-DNN} performs source separation directly in the wave domain without back-end electronic processing, successfully demixing multiple mixed signals sharing identical time-frequency characteristics and polarization states. The current system demonstrates $10$ separation channels, with the potential for further scaling to increase channel capacity.




\clearpage
\newpage

\begin{figure}[t!]
	\centering
	\includegraphics[width=1\textwidth]{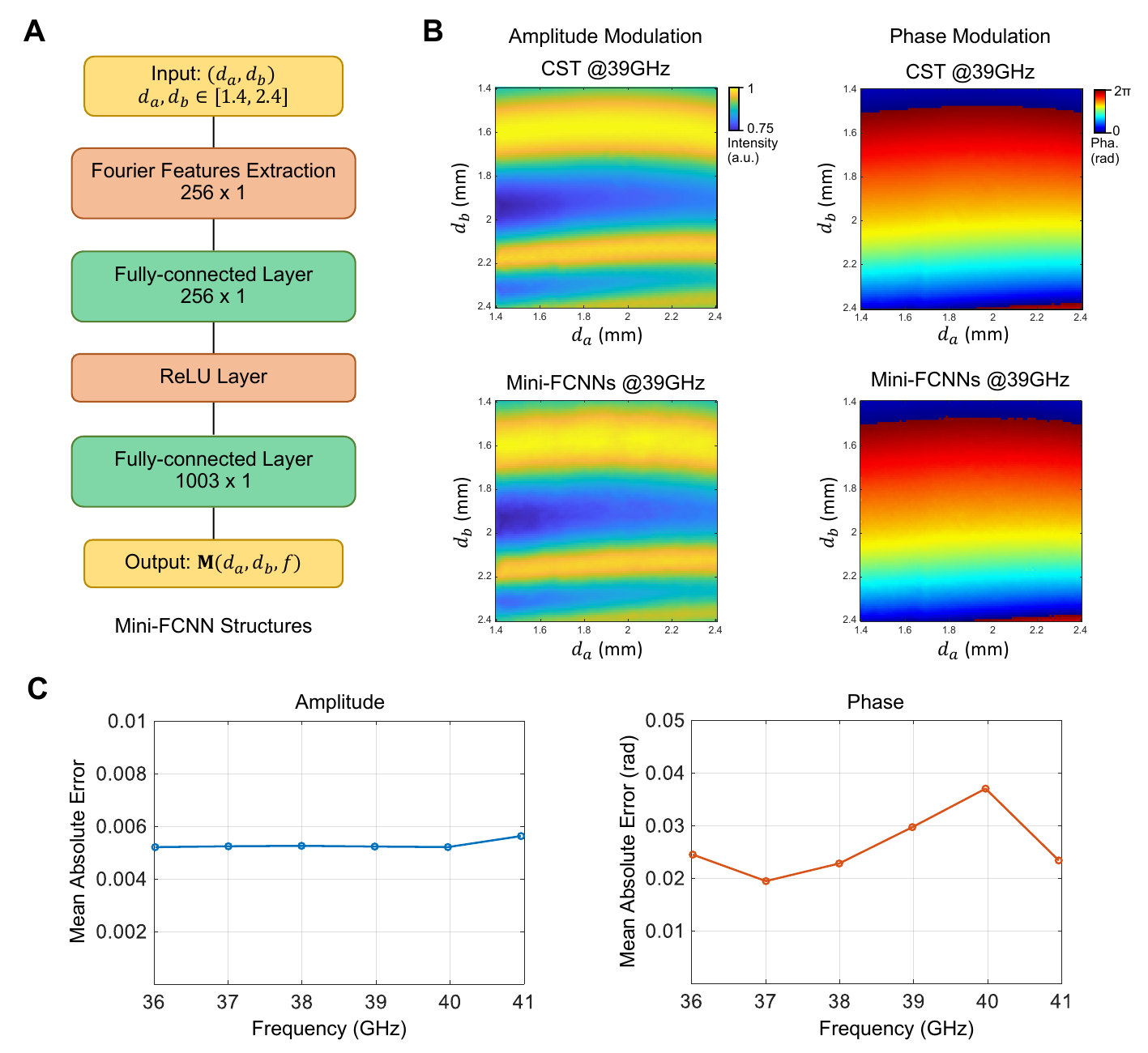}
	\caption{\textbf{Mini-FCNN for predicting meta-atom modulations.} (\textbf{A}) The mini-FCNN structure comprises a Fourier feature extraction layer and two fully-connected layers followed by ReLU activation. (\textbf{B}) Comparison between mini-FCNN predictions and CST simulations for meta-atom amplitude and phase responses. The high similarity between the modulation values across different geometric settings demonstrates the effectiveness of the proposed neural network for multi-dimensional modulation prediction. (\textbf{C}) The mean absolute errors of amplitude and phase modulation predictions under different frequencies using mini-FCNN.}
	\label{Mini-FCNN_Structure}
\end{figure}

\clearpage
\newpage

\begin{figure}[t!]
	\centering
	\includegraphics[width=1\textwidth]{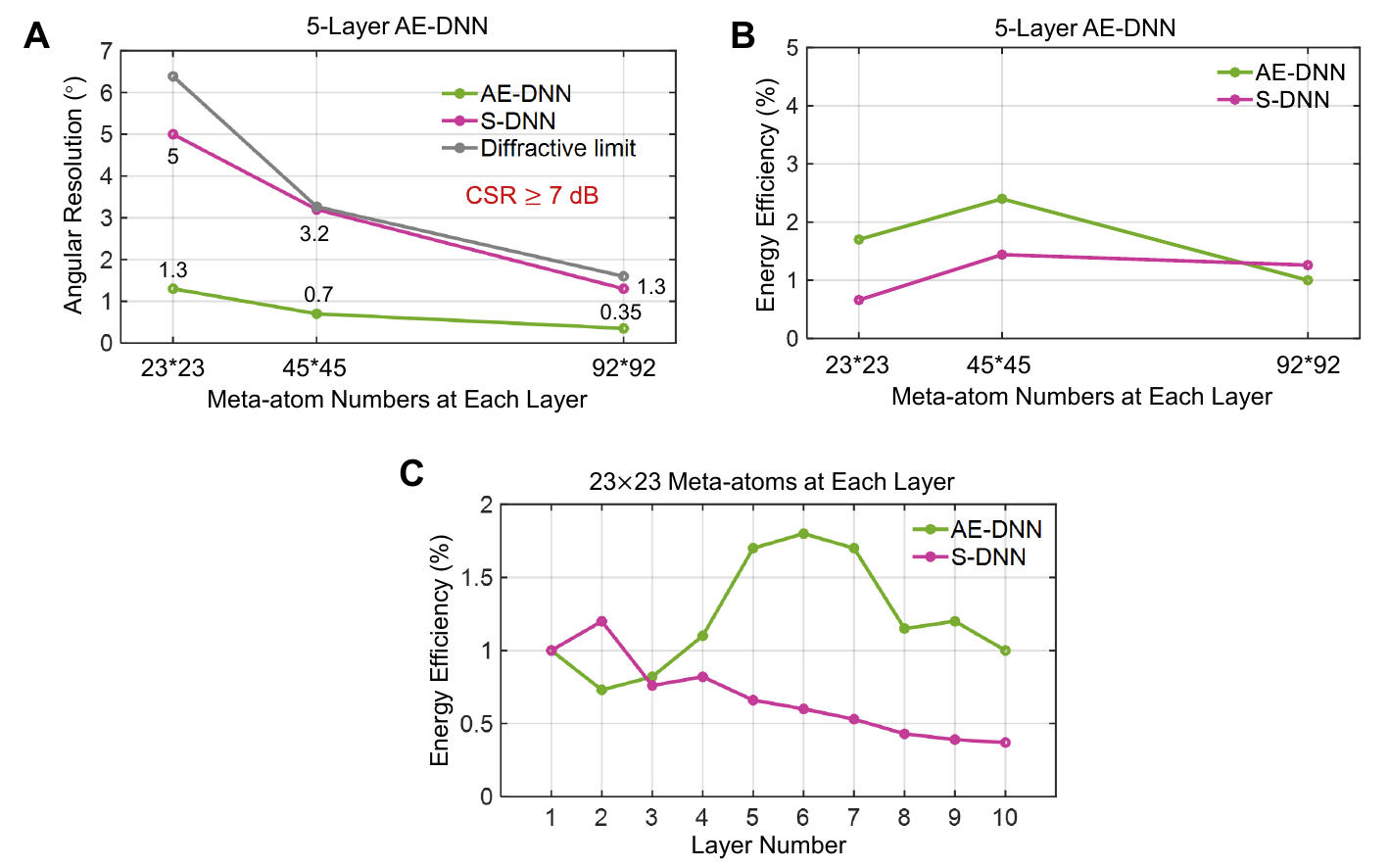} 
	\caption{\textbf{AE-DNN performance under varying network settings.} (\textbf{A}) Relationship between AE-DNN angular resolution and aperture size.  (\textbf{B}) Relationship between AE-DNN energy efficiency and aperture size. (\textbf{C}) Relationship between AE-DNN energy efficiency and number of layers. The energy efficiency of AE-DNN does not decrease significantly with increasing layer number and propagation distance.}
	\label{Ablation_Study}
\end{figure}

\clearpage
\newpage

\begin{figure}[t!]
	\centering
	\includegraphics[width=1\textwidth]{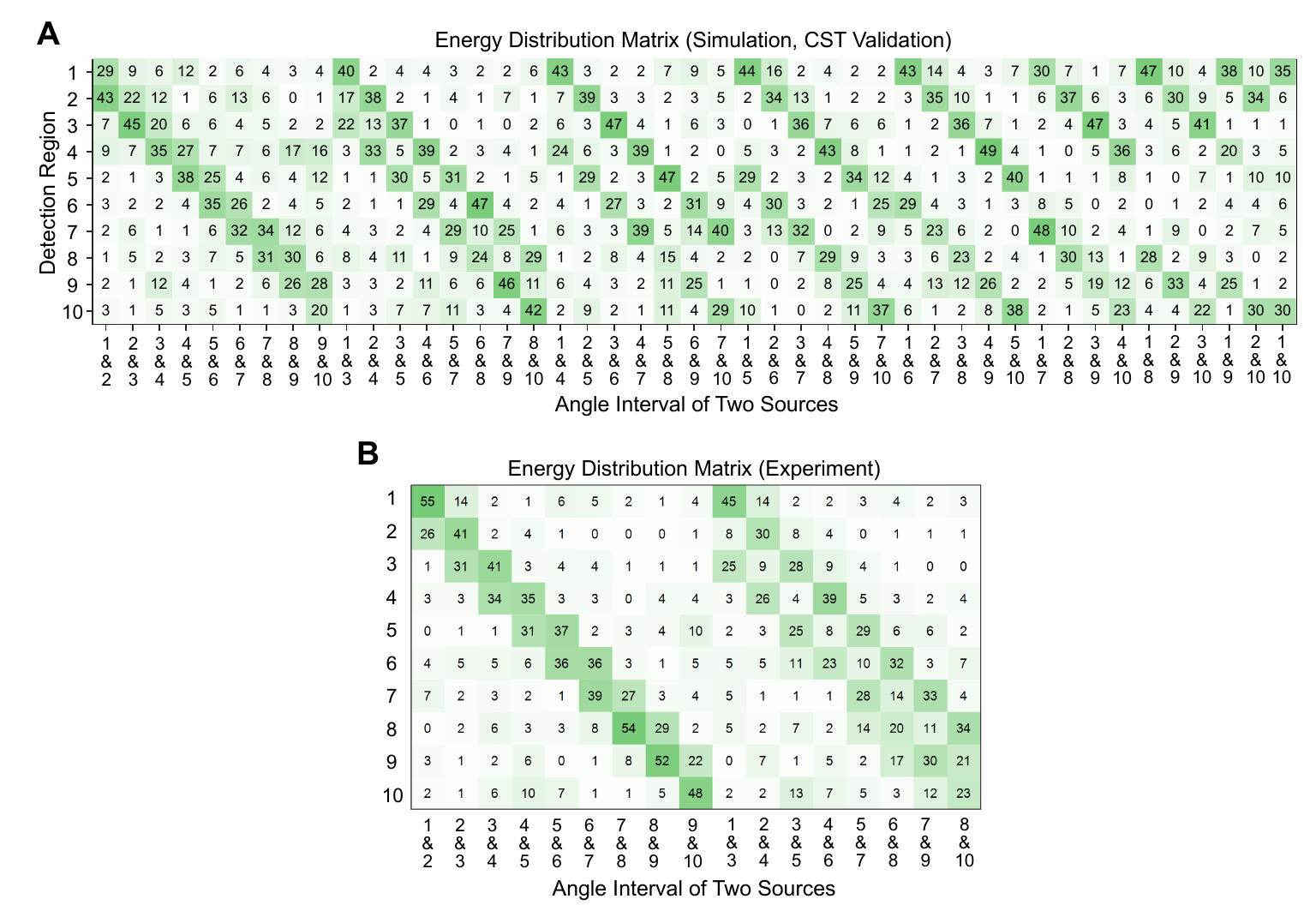}
	\caption{\textbf{Super-resolution DOA estimation results for 5-layer AE-DNN.} (\textbf{A}) CST simulation energy distribution matrix of AE-DNN performing 1.6° super-resolution DOA estimation for 45 two-source angular permutations. (\textbf{B}) Experimental energy matrix of AE-DNN performing 1.6° super-resolution DOA estimation.}
	\label{Five-layer_AE-DNN_Energy_Matrix}
\end{figure}

\clearpage
\newpage

\begin{figure}[t!]
	\centering
	\includegraphics[width=1\textwidth]{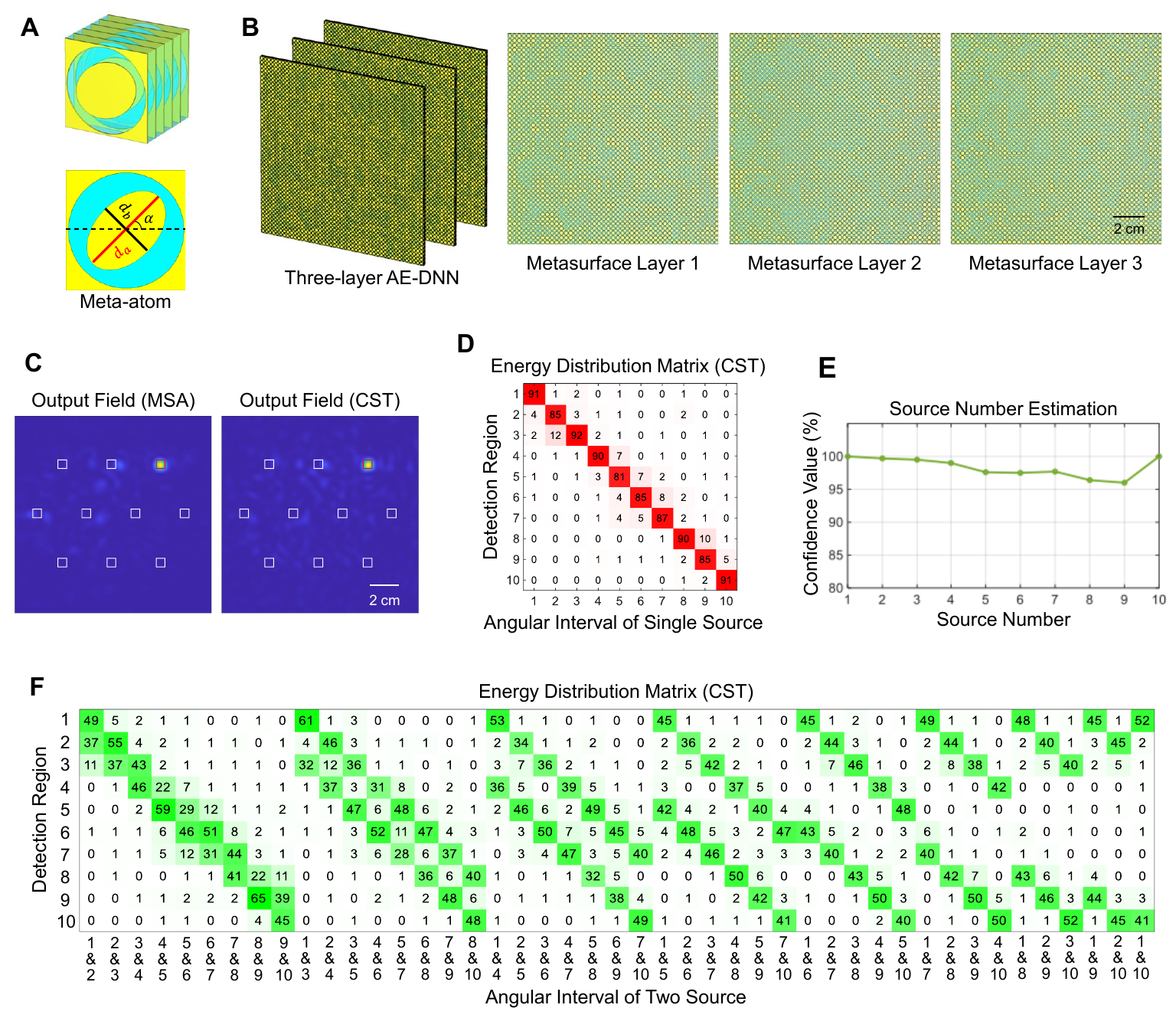}
	\caption{\textbf{Numerical evaluations of three-layer AE-DNN.} (\textbf{A}) Schematic diagram of each meta-atom of the metasurface. The geometric parameters $\left(d_a,\, d_b,\, \alpha\right)$ affect the EM wave modulation effect. (\textbf{B}) AE-DNN's CST model and the structure of each metasurface. (\textbf{C}) Example inference results for a single source (angle of -7.5°) using both MSA simulation and CST evaluations. (\textbf{D}) CST energy distribution matrix evaluated on the single-source testing dataset. (\textbf{E}) AE-DNN can perform source number estimation tasks on up to 10 sources. (\textbf{F}) CST energy distribution matrix of AE-DNN performing 3° super-resolution DOA estimation of two sources under 45 angular interval permutations.}
	\label{Three-layer_AE-DNN_Evaluation}
\end{figure}

\clearpage
\newpage

\begin{figure}[t!]
	\centering
	\includegraphics[width=1\textwidth]{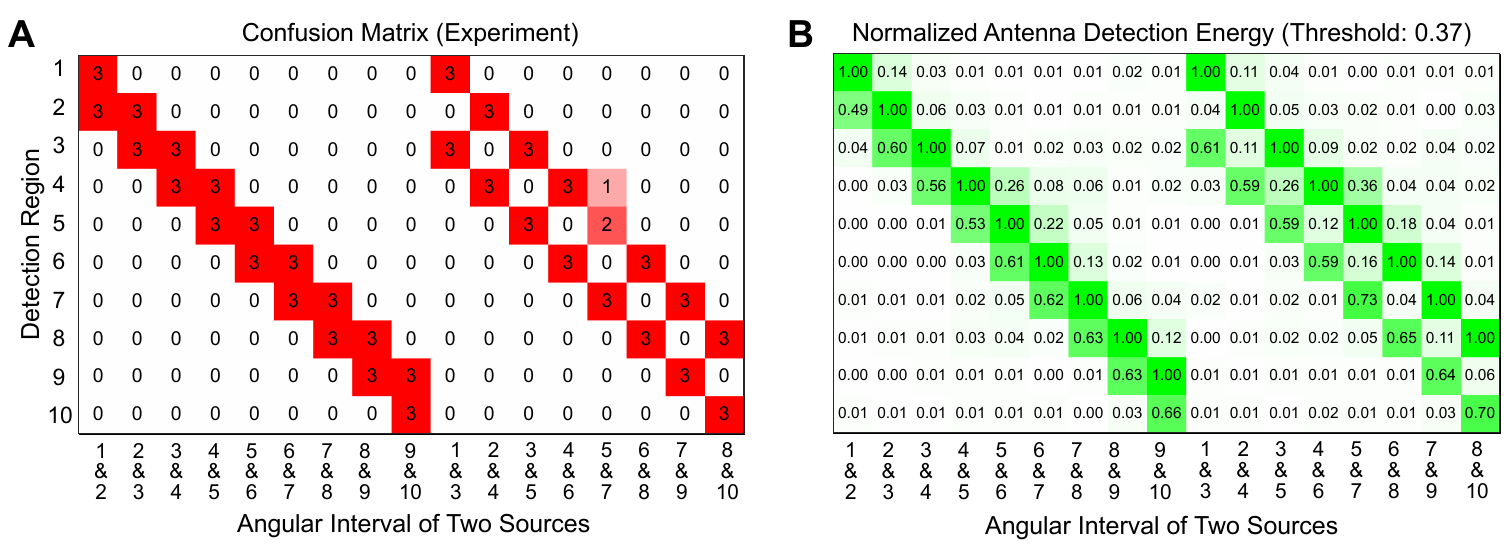}
    \caption{\textbf{Experimental performance of AE-DNN for source number estimation and super-resolution DOA estimation in ISCC applications.} (\textbf{A}) The confusion matrix of the DOA estimation task was evaluated on 51 sets of two-source experimental samples, with the angular interval between the BPSK and QAM16 being 3°or 6°. The confidence value of the DOA estimation task was $98\%$.  (\textbf{B}) The calibrated and normalized antenna detection energy for the source number estimation task was measured on the same experimental samples, with a threshold of 0.37 and a confidence value of $96\%$.}
	\label{Ten-Source_Separation_Experiment}
\end{figure}
\clearpage
\newpage

\begin{figure}[t!]
	\centering
	\includegraphics[width=1\textwidth]{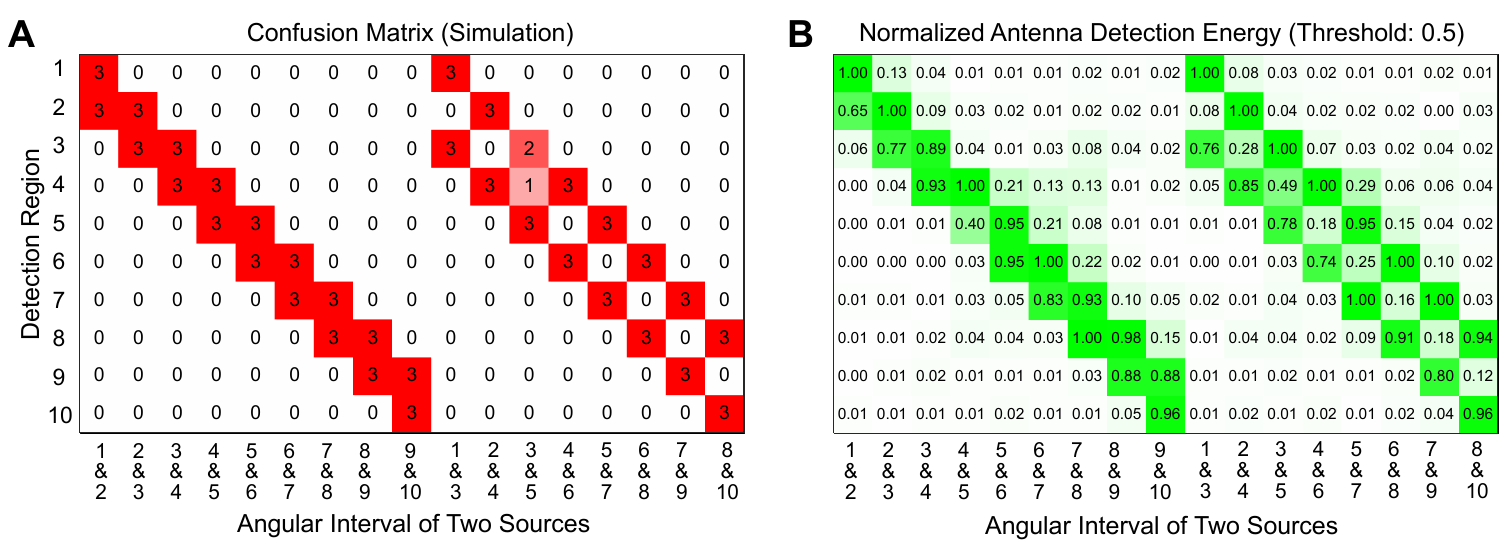}
	\caption{\textbf{Experimental performance of AE-DNN for source number estimation and super-resolution DOA estimation in radar anti-jamming applications.} (\textbf{A}) The confusion matrix of the DOA estimation task is evaluated on 51 sets of two-source experimental samples, with the angular interval between the radar signal and the jamming being 3°or 6°. The confidence value of the DOA estimation task was $98\%$.  (\textbf{B}) The calibrated and normalized antenna detection energy for the source number estimation task was measured on the same experimental samples, with a threshold of 0.5 and a confidence value of $94.2\%$.}
	\label{Ten-Source_Separation_Simulation}
\end{figure}

\clearpage
\newpage

\begin{figure}[t!]
	\centering
	\includegraphics[width=1\textwidth]{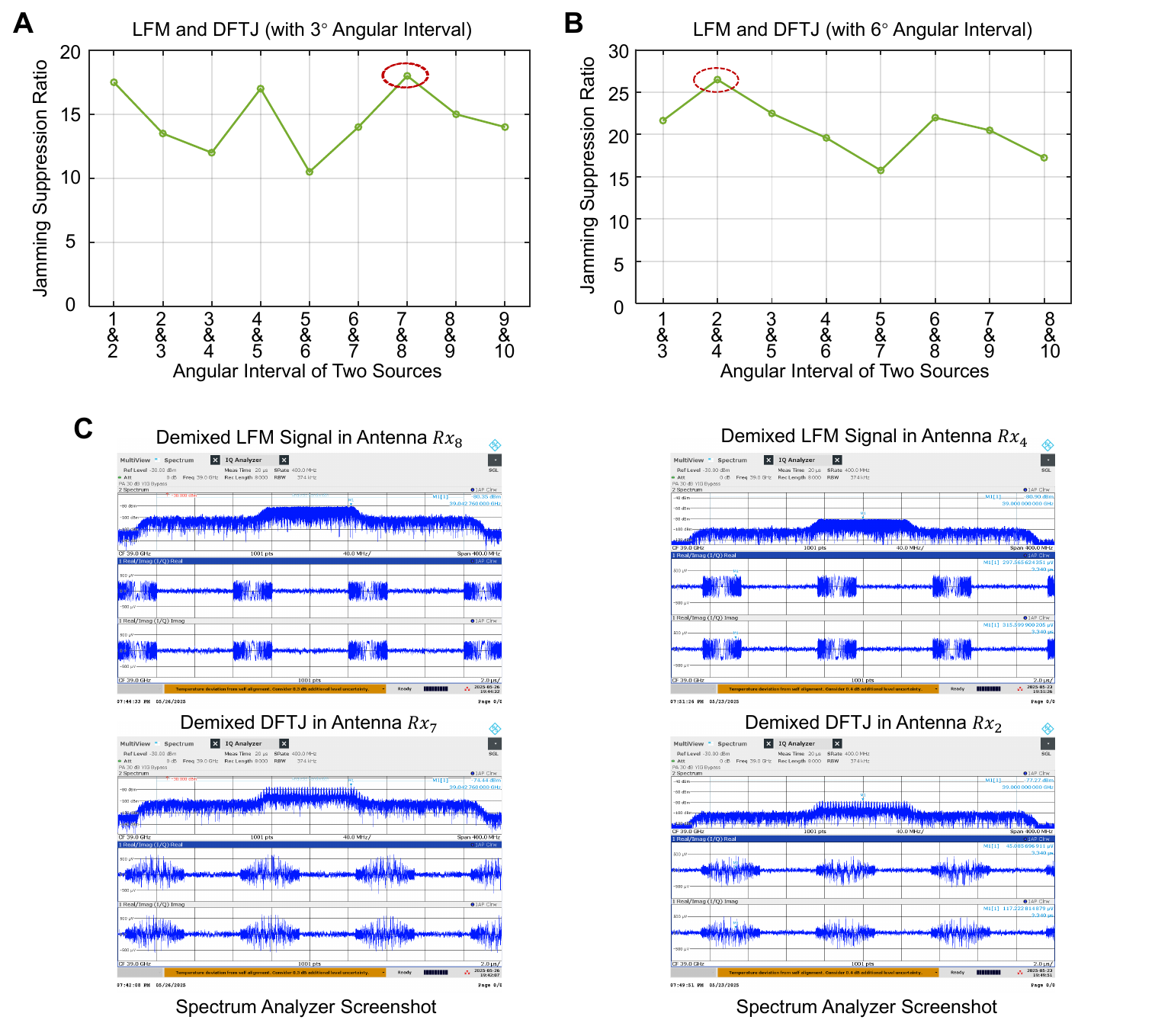}
	\caption{\textbf{Experimental performance of AE-DNN for source separation in radar anti-jamming applications.} (\textbf{A}) The jamming suppression ratio (JSR) of source separation was evaluated on multiple sets of two-source experimental samples. When the angular separation between the LFM signal and the DFTJ was 3°, the average JSR was 14.6 dB.  (\textbf{B}) When the angular separation between the LFM signal and the DFTJ was 6°, the average JSR was 20 dB. (\textbf{C}) Below are screenshots of the spectrum analyzer for different demixing channels.}
	\label{Radar_Anti-jamming_Screenshot}
\end{figure}

\clearpage
\newpage

\begin{figure}[t!]
	\centering
	\includegraphics[width=1\textwidth]{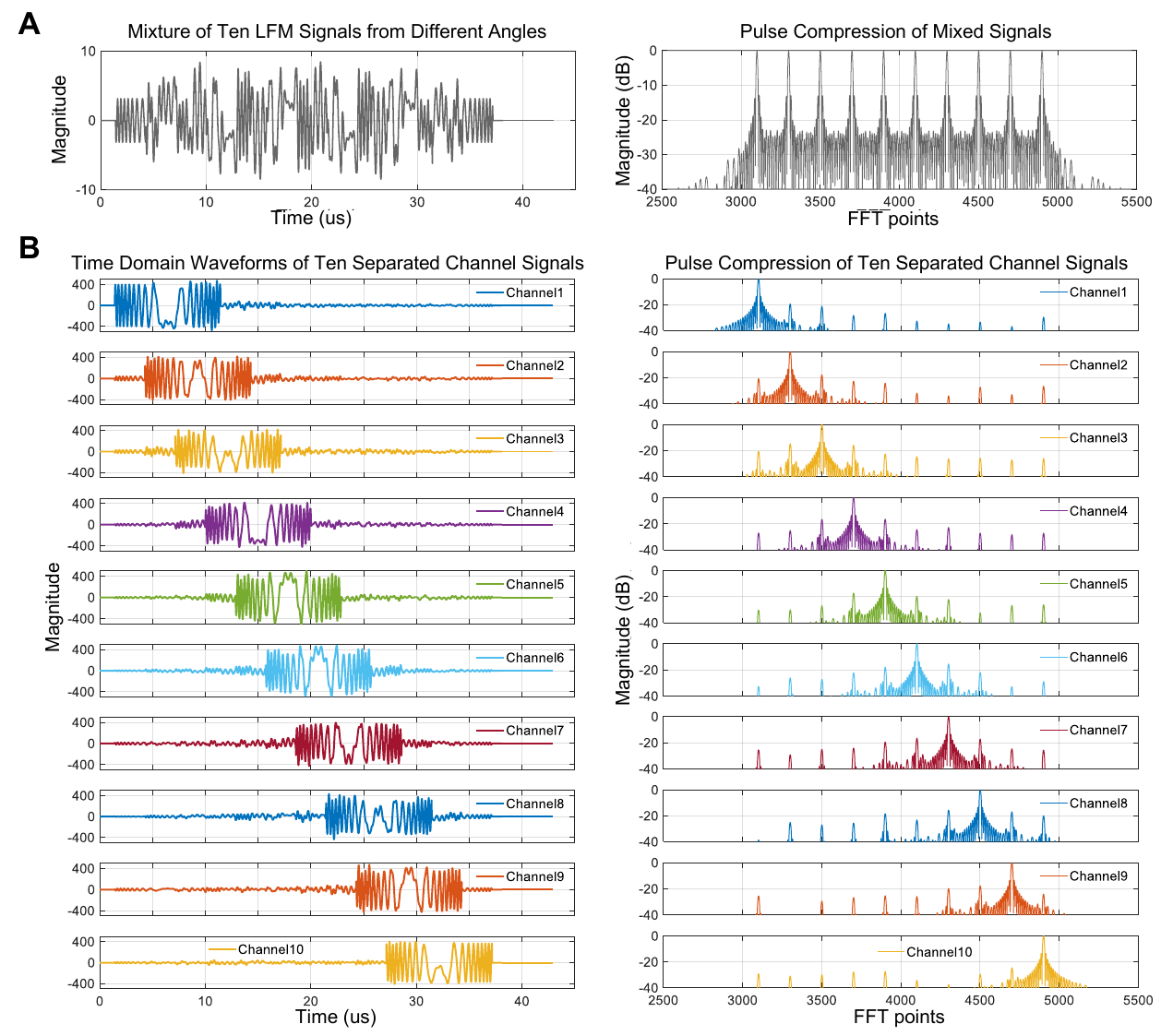}
	\caption{\textbf{Simulation proves that AE-DNN supports up to 10 source separation channels.} (\textbf{A}) The simulation generates ten LFM signals with different angles and time delays, and the time domain waveform and pulse compression of the mixed signal after coherent superposition. (\textbf{B}) After AE-DNN source separation, the waveforms and pulse compression results of the demixed signals in the ten separated channels are shown. The average channel crosstalk suppression ratio is 17.6 dB.}
	\label{Ten-Source_Separation_ISCC}
\end{figure}

\clearpage
\newpage

\begin{figure}[t!]
	\centering
	\includegraphics[width=1\textwidth]{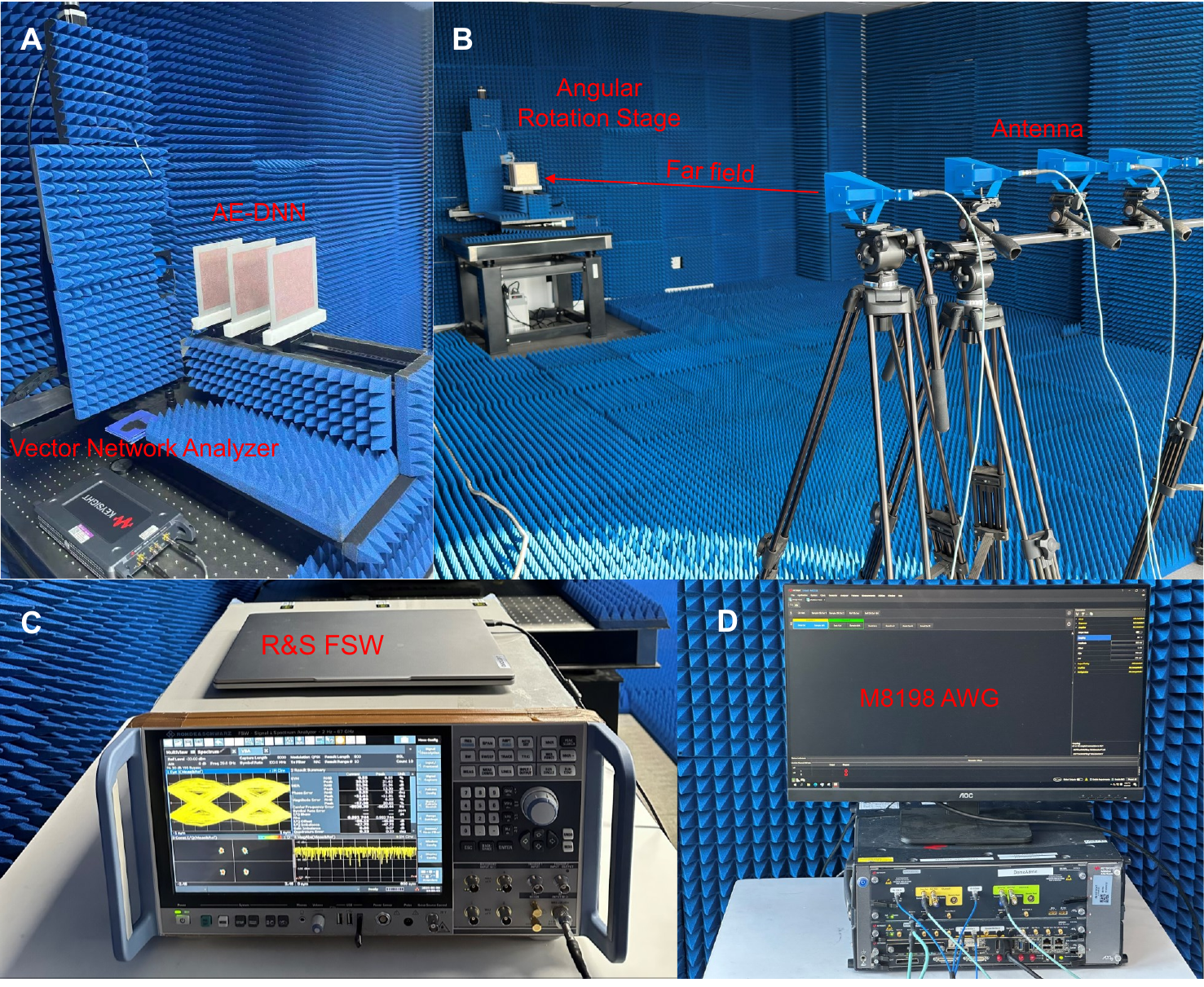}
	\caption{\textbf{Experimental setup for testing AE-DNN.} (\textbf{A}) Overview of the microwave anechoic chamber, including the vector network analyzer (VNA), the AE-DNN fixed on an angle rotating stage, and waveguide probe mounted on an XY translation stage. (\textbf{B}) Multiple transmitting horn antennas are located at the far field distance of the AE-DNN. (\textbf{C}) The R\&S spectrum analyzer for demodulating and analyzing radar and communication signals. (\textbf{D}) Keysight Technologies' M8198 arbitrary waveform generator, capable of simultaneously transmitting two signals, including radar signals and jamming, or communication signals with different modulation formats.}
	\label{Experimental_Setup}
\end{figure}

\clearpage
\newpage

\begin{table} 
	\centering
	\caption{\textbf{Output Energy Response Vector of AE-DNN (Experiment)}}
	\label{tab_S2} 

	\begin{tabular}{lccr} 
		\\
		\hline
		\textbf{Incident Angle} & \textbf{Normalized Output Energy Response Vector (Unit: a.u.)} \\
		\hline
        $\theta_{1}=-13.5^\circ$ & 
        $\mathbf{P}_y=
        \left[\mathbf{1},\,0.01,\,0.01,\,0.01,\,0.01,\,0.01,\,0.01,\,0.01,\,0.01,\,0.01\right]$ \\
        
        $\theta_{2}=-10.5^\circ$ & 
        $\mathbf{P}_y=
        \left[0.1,\,\mathbf{0.74},\,0.05,\,0.01,\,0.01,\,0.01,\,0.01,\,0.01,\,0.01,\,0.01\right]$ \\
        
        $\theta_{3}=-7.5^\circ$ & 
        $\mathbf{P}_y=
        \left[0.02,\,0.03,\,\mathbf{0.36},\,0.01,\,0.01,\,0.01,\,0.01,\,0.01,\,0.01,\,0.01\right]$ \\
        
        $\theta_{4}=-4.5^\circ$ & 
        $\mathbf{P}_y=
        \left[0.01,\,0.01,\,0.01,\,\mathbf{0.39},\,0.01,\,0.01,\,0.01,\,0.01,\,0.01,\,0.01\right]$ \\
        
        $\theta_{5}=-1.5^\circ$ & 
        $\mathbf{P}_y=
        \left[0.01,\,0.01,\,0.01,\,0.05,\,\mathbf{0.64},\,0.01,\,0.01,\,0.01,\,0.01,\,0.01\right]$ \\
        
        $\theta_{6}=1.5^\circ$ & 
        $\mathbf{P}_y=
        \left[0.01,\,0.01,\,0.01,\,0.01,\,0.01,\,\mathbf{0.63},\,0.02,\,0.01,\,0.01,\,0.01\right]$ \\
        
        $\theta_{7}=4.5^\circ$ & 
        $\mathbf{P}_y=
        \left[0.01,\,0.01,\,0.01,\,0.01,\,0.02,\,0.04,\,\mathbf{0.57},\,0.02,\,0.01,\,0.01\right]$ \\
        
        $\theta_{8}=7.5^\circ$ & 
        $\mathbf{P}_y=
        \left[0.01,\,0.01,\,0.01,\,0.01,\,0.01,\,0.01,\,0.01,\,\mathbf{0.5},\,0.01,\,0.01\right]$ \\
        
        $\theta_{9}=10.5^\circ$ & 
        $\mathbf{P}_y=
        \left[0.01,\,0.01,\,0.01,\,0.01,\,0.02,\,0.01,\,0.01,\,0.07,\,\mathbf{0.67},\,0.01\right]$ \\
        
        $\theta_{10}=13.5^\circ$ & 
        $\mathbf{P}_y=
        \left[0.01,\,0.01,\,0.01,\,0.01,\,0.01,\,0.01,\,0.01,\,0.03,\,0.07,\,\mathbf{0.68}\right]$ \\
        \hline
	\end{tabular}
\end{table}

\begin{table} 
	\centering
	\caption{\textbf{Output Energy Response Vector of AE-DNN (Simulation)}}
	\label{tab_S1} 

	\begin{tabular}{lccr} 
		\\
		\hline
		\textbf{Incident Angle} & \textbf{Normalized Output Energy Response Vector (Unit: a.u.)} \\
		\hline
        $\theta_{1}=-13.5^{\circ}$ & $\mathbf{P}_y=\left[\mathbf{1},0.02,0.01,0.01,0.01,0.01,0.01,0.01,0.01,0.01\right]$ \\
        $\theta_{2}=-10.5^{\circ}$ & $\mathbf{P}_y=\left[0.02,\mathbf{0.88},0.03,0.01,0.01,0.01,0.01,0.01,0.01,0.01\right]$ \\
        $\theta_{3}=-7.5^{\circ}$  & $\mathbf{P}_y=\left[0.01,0.02,\mathbf{0.7},0.02,0.01,0.01,0.01,0.01,0.01,0.01\right]$ \\
        $\theta_{4}=-4.5^{\circ}$  & $\mathbf{P}_y=\left[0.01,0.01,0.02,\mathbf{0.76},0.03,0.01,0.01,0.01,0.01,0.01\right]$ \\
        $\theta_{5}=-1.5^{\circ}$  & $\mathbf{P}_y=\left[0.01,0.01,0.01,0.02,\mathbf{0.98},0.03,0.01,0.01,0.01,0.01\right]$ \\
        $\theta_{6}=1.5^{\circ}$   & $\mathbf{P}_y=\left[0.01,0.01,0.01,0.01,0.02,\mathbf{0.9},0.02,0.01,0.01,0.01\right]$ \\
        $\theta_{7}=4.5^{\circ}$   & $\mathbf{P}_y=\left[0.01,0.01,0.01,0.01,0.01,0.04,\mathbf{0.78},0.02,0.01,0.01\right]$ \\
        $\theta_{8}=7.5^{\circ}$   & $\mathbf{P}_y=\left[0.01,0.01,0.01,0.01,0.01,0.01,0.02,\mathbf{0.7},0.03,0.01\right]$ \\
        $\theta_{9}=10.5^{\circ}$  & $\mathbf{P}_y=\left[0.01,0.01,0.01,0.01,0.01,0.01,0.01,0.02,\mathbf{0.9},0.01\right]$ \\
        $\theta_{10}=13.5^{\circ}$ & $\mathbf{P}_y=\left[0.01,0.01,0.01,0.01,0.01,0.01,0.01,0.01,0.02,\mathbf{0.85}\right]$ \\
        \hline
	\end{tabular}
\end{table}



\end{document}